# A comparison of different Fourier transform procedures for analysis of diffraction data from noble gas fluids

J.E. Proctor[1], C.G. Pruteanu[2], B. Moss[1], M.A. Kuzovnikov[2], G.J. Ackland[2], C.W. Monk[1] and S. Anzellini[3]

1. Materials and Physics Research Group, School of Science, Engineering and Environment, University of Salford, Manchester M5 4WT, UK

2. SUPA, School of Physics & Astronomy and Centre for Science at Extreme Conditions, the University of Edinburgh, Edinburgh EH9 3FD, UK

3. Diamond Light Source Ltd., Harwell Science and Innovation Campus, Diamond House, Didcot OX11 0DE, UK

**Abstract**

A comparison is made between the three principal methods for analysis of neutron and X-ray diffraction data from noble gas fluids by direct Fourier transform. All three methods (standard Fourier transform, Lorch modification and Soper-Barney modification) are used to analyse four different sets of diffraction data from noble gas fluids. The results are compared to the findings of a full-scale real space structure determination, namely Empirical Potential Structure Refinement. Conclusions are drawn on the relative merits of the three Fourier transform methods, what information can be reliably obtained using each method, and which method is most suitable for analysis of different kinds of diffraction data. The mathematical validity of the Lorch method is critically analysed.

## Introduction

### I. Analysis of fluid diffraction data via Fourier transform: The $Q_{max}$ cutoff problem

On a fundamental level, the interpretation of diffraction data from fluids, network glasses and amorphous solids is simple. The coherent scattering intensity $I_{coh}(Q)$ (of X-rays or neutrons) is predicted by the Debye scattering equation, given below for a fluid composed of spherically symmetric identical particles:

$$I_{coh}(Q) = \sum_{m=1}^{N} \left[ \sum_{n=1}^{N} \frac{f(Q)^2 \sin(Qr_{mn})}{Qr_{mn}} \right]$$

$$(1)$$

Here $f(Q)$ is the atomic form factor (X-rays) or scattering length (neutrons) (tabulated data are available for this [1]) and $N$ is the number of particles in the beam. The parameter $r_{mn}$ refers to the distance between particles $m$ and $n$. The structure factor $S(Q)$ is obtained from the coherent scattering intensity $I_{coh}(Q)$ according to equation 2 [2][3]:

$$S(Q) = \frac{I_{coh}(Q)}{Nf(Q)^2}$$

(2)

If the coherent scattering intensity $I_{coh}(Q)$ is defined according to the Debye scattering equation, then the definition of $S(Q)$ according to equation 1 will result in $\lim_{Q \to \infty} S(Q) = 1$.

In reality $I_{coh}(Q)$, as defined by the Debye scattering equation, cannot be obtained. Instead the experimentally measured scattering data undergoes a series of corrections to calculate the raw coherent scattering intensity from the sample $I_{raw}(Q)$ as reliably as possible. The remaining normalization to obtain $S(Q)$ is performed using equation 3 through division by $N'f(Q)^2$, where $N'$ is an arbitrary scaling parameter to ensure that ensuring that $\lim_{Q \to \infty} S(Q) = 1$. The normalization via $N'$ accounts for several effects including the fact that $I_{raw}(Q)$ (unlike $I_{coh}(Q)$) depends on the incident beam intensity.

$$S(Q) = \frac{I_{raw}(Q)}{N'f(Q)^2}$$

(3)

All further analysis utilizes this definition of $S(Q)$.

The radial distribution function $g(r)$ enables parameters such as the co-ordination number (CN) describing sample properties in real space to be obtained. The $g(r)$ function can be derived by a Fourier transform (FT) of the structure factor $S(Q)$ provided by the diffraction data. For a fluid composed of spherically symmetric identical particles the FT relationship given below in equation 4 [2][3] is exactly correct. The parameter $g_0$ is an arbitrary constant which is eliminated in any case when $g(r)$ is normalized (as outlined in the supplementary information).

$$4\pi r^2 [g(r) - g_0] = \frac{2}{\pi} \int_0^\infty Qr[S(Q) - 1]\sin(Qr)\,dQ$$

(4)

By definition, $g(r)$ represents the probability of finding particles separated by a particular distance. Therefore $g(r)$ has certain *mathematical* properties which most arbitrary functions do not. Notably, only functions which can be derived from an ensemble of point positions in space are possible $g(r)$ functions.

For fluids there are further *physical* constraints: In the absence of long range order one must have $g(r) = 1$ at long range, furthermore $g(r)$ and its derivatives must be continuous. From equation 4, these constraints also apply to $S(Q)$.

We can therefore identify four distinct constraints on any meaningful $S(Q)$:

1. The associated $g(r)$ must correspond to an ensemble of points in space.
2. For atoms with finite size, $g(r)$ must be positive everywhere and go to zero for $r$ approximately equal to the atomic diameter.
3. $S(Q)$ must be continuous.
4. Derivatives of $S(Q)$ must be continuous.

In fact, constraints 3 and 4 are consequences of constraint 1, but we list them separately because methods exist to resolve them. Only a very small fraction of possible functions satisfy these necessary constraints, and there is no reason to suppose that $S(Q)$ derived from real data via equation 3 will do so.

One problem is that real diffraction data cover a range in $Q$ only from a finite minimum value of $Q$, $Q_{min}$, up to some finite value, $Q_{max}$. In this work we will focus primarily on solutions to the problems caused by $Q_{max} < \infty$. We will however return to a discussion of the potential problems caused by $Q_{min} > 0$ in the conclusions. The $Q_{max}$ problem is particularly severe for X-ray diffraction, where destructive interference between X-rays scattered from different parts of the same atom prevents scattering at large $Q$. This phenomenon is represented mathematically using the parameter known as the atomic form factor $f(Q)$. Even for neutron diffraction, the range in $Q$ covered by the data is limited by its relation to the scattering angle (which cannot exceed 180°), and by the reduction in $I_{coh}(Q)$ at large $Q$ that takes place independently of the effect of changes to $f(Q)$, via the $1/Q$ factor in the Debye scattering equation (equation 1). This phenomenon, in both X-ray and neutron diffraction experiments, is known as the $Q_{max}$-cutoff problem. This problem leads to all four types of unphysicalities described above.

In the absence of data, Equation 4 can only be applied to experimental data if some theoretical assumption is made about $S(Q > Q_{max})$. The simplest theory is to take $[S(Q > Q_{max}) - 1] = 0$ (equivalently, integrate only to $Q_{max}$). This "absence of evidence is evidence of absence" theory is so widely used that it is often not even recognised as an assumption. It almost certainly violates all the physical constraints described above: most obviously it introduces a discontinuity in S(Q).

Mathematically, this sharp cutoff at $Q_{max}$ means that the FT generates a "$g(r)$" which is a convolution of the true $g(r)$ data with the FT of a step function $f_S(Q)$: this latter is a sinc function which produces spurious oscillations at a frequency determined by $Q_{max}$. These manifest as peaks in $g(r)$ at unfeasibly small $r$, and in extreme cases, even negative $g(r)$ (both leading to $g(r)$ violating condition 2 above).

We can write the FT of data up to $Q_{max}$ as:

$$4\pi r^2 [g(r) - g_0] = \frac{2}{\pi} \int_0^\infty Qr[S(Q) - 1] \sin(Qr) f_S(Q) dQ$$
$$= \frac{2}{\pi} \int_0^{Q_{max}} Qr[S(Q) - 1] \sin(Qr) f_S(Q) dQ$$

(5)

Where $f_S(Q)$ is the step function (See figure 1 later), which means that beyond $Q_{max}$, the integrand is zero and can be ignored. We note that if the data are extensive enough that $[S(Q > Q_{max}) - 1] = 0$, this also gives a zero integrand above $Q_{max}$ and the step function is unnecessary.

This discussion may appear unnecessarily pedantic, but its purpose is to introduce the concept of $f_S(Q)$. While the step function preserves the integrity of the data up to $Q_{max}$, *any* function that is equal to zero above $Q_{max}$ makes it possible to perform a complete FT in spite of the missing data. As described, the step function (sinc convolution) may produce nonsensical results which is typically undesirable. A different function, which modifies the data but does not introduce a discontinuity, may produce a $g(r)$ which is less obviously wrong.

## II. Solution to the $Q_{max}$ cutoff problem by the Lorch modification function

In 1969, Lorch proposed a solution [4] to the discontinuity arising from the $Q_{max}$ cutoff problem, beginning from equation 4. Where we use $S(Q)$, Lorch used the notation $i(Q)$. Lorch's method is essentially to make a compromise. Instead of seeking to determine the exact value of $g(r)$ for all $r$, we will obtain the integrated value of $g(r)$ over a range of values about some central value $r_0$: $(r_0 - \Delta) \leq r \leq (r_0 + \Delta)$ (Lorch used $\Delta/2$ instead). In this case, both sides of equation 4 are integrated with respect to $r$ within these limits and Lorch obtained an equation which is reproduced below using our notation as equation 6.

$$4\pi r_0^2 [g(r_0) - g_0] = \frac{2}{\pi} \int_0^{Q_{max}} Q r_0 [S(Q) - 1] \sin(Q r_0) \left[ \frac{\sin(Q\Delta)}{Q\Delta} \right] dQ$$

$$f_L(Q\Delta) = \frac{\sin(Q\Delta)}{Q\Delta}$$

$$(6)$$

The result of this compromise is an expression on the right hand side identical to equation 5 except that the integrand is now multiplied by a modification function $f_L(Q\Delta)$, the sinc function. $Q_{max} = \pi/\Delta$ is usually chosen as this makes $Q_{max}\Delta$ correspond to the first zero in the sinc function. In this case constraint 3 as listed above will be satisfied: We now integrate $[S(Q) - 1]f_L$ which is continuous at $Q_{max}$ instead of $[S(Q) - 1]f_S$, which is not. Beyond $Q_{max}$, the integrand of the FT must be zero: this can be achieved by some assumption on the data: $[S(Q > Q_{max}) - 1] = 0$, or by setting $f_L(Q > Q_{max}) = 0$. These yield identical results: to maintain compatibility with the step function approach, we adopt the second interpretation throughout.

Equation 6 has been used as a standard method to address the $Q_{max}$-cutoff discontinuity problem for the 50 years since Lorch's paper was published; it has 571 citations to date, 125 of which are from the last 5 years. The method has been used in many papers fundamental to our understanding of fluids, network glasses and amorphous solids.

However, in 2011 [5] it was shown that, if the integral is performed as instructed in Lorch's original paper one in fact obtains the equation reproduced below in our notation as equation 7. Here, the application of a cutoff at some finite value of $Q$ cannot be justified for the second integral, which is not necessarily small compared to the first integral.

$$4\pi r_0^2 [g(r_0) - g_0]$$
$$= \frac{2}{\pi} \int_0^{Q_{max}} Q r_0 [S(Q) - 1] \sin(Q r_0) \left[ \frac{\sin(Q\Delta)}{Q\Delta} \right] dQ$$
$$+ \frac{2}{\pi} \int_0^{\infty} [S(Q) - 1] \left[ \frac{\sin(Q\Delta)\cos(Q r_0)}{Q\Delta} - \cos(Q r_0)\cos(Q\Delta) \right] dQ$$

$$(7)$$

Over the decades, Lorch's method has been very successful at removing unphysical oscillations in $g(r)$ functions but this does not prove that the method is producing $g(r)$ functions which are correct. This naturally raises the question: can Lorch's result (equation 6) be recovered through the use of reasonable approximations? Let us examine the problem carefully.

To begin, we integrate equation 3 exactly as described within the specified limits (equation 8). The left hand side gives us the integrated value of $g(r)$ straightforwardly but, in order to continue referring to the non-integrated value of $g(r)$ (as is done in ref. [4] and in equations 6 and 7) it is necessary to make additional assumptions. We assume that $\Delta \ll r_0$ and that $g(r)$ does not vary significantly over the region covered by the integral over $r$. Equation 9 below specifies the exact, and approximate, relationships between $g(r)$ and its integrated equivalent which we shall label $G(r_0, \Delta)$.

$$\int_{r_0-\Delta}^{r_0+\Delta} 4\pi r_0^2 [g(r) - g_0] dr = \frac{2}{\pi} \iint_{r=r_0-\Delta, Q=0}^{r=r_0+\Delta, Q=\infty} Qr[S(Q) - 1]\sin(Qr)\, dQ dr$$

(8)

$$G(r_0, \Delta) = \int_{r_0-\Delta}^{r_0+\Delta} 4\pi r_0^2 [g(r) - g_0] dr$$

$$G(r_0, \Delta) \approx 2\Delta \times 4\pi r_0^2 [g(r_0) - g_0]$$

(9)

The right hand side of equation 8 is a standard integral of the form $\int x \sin x\, dx$. Application of trigonometric identities $\cos(A + B) = \cos A \cos B - \sin A \sin B$ etc. results in equation 7, in agreement with ref. [5].

Nevertheless, we will now see that the Lorch function is recoverable. If we implement the condition that $\Delta \ll r_0$ on the right hand side of equation 8 also, then the variation of the $r$ term within the integrand is negligible over the range of the integral so we can replace it with $r_0$ and simply integrate $\sin(Qr)$ with respect to $r$. In this case, again by making use of standard trigonometric identities, equation 4 is recovered, i.e. the Lorch function is mathematically valid under this condition.

It is not correct to also set $\sin(Qr) \to \sin(Qr_0)$ in equation 8, as it would necessitate the far more stringent condition on $\Delta$ that $\Delta \to 0$. This is because we would require the variation in the value of $Qr$ over the course of $r_0 - \Delta \le r \le r_0 + \Delta$ to cause negligible change to the value of $\sin(Qr)$ even in the high $Q$ limit.

## III. Other solutions to the discontinuity at $Q_{max}$ cutoff problem

Since the work of Lorch, several other solutions have been proposed to the $Q_{max}$ cutoff problem. The most notable one is the modification function proposed by Soper and Barney [5]. Similarly to the Lorch function, this function involves accepting an averaged value of $g(r)$ over a range $\pm\Delta$. This function (equation 10) takes values in the low $Q$ and high $Q$ limits identical to the sinc function proposed by Lorch (equation 6) and is inserted into the FT equation in an analogous manner (equation 11).

$$f_{SB}(Q\Delta) = \frac{3}{(Q\Delta)^3}(\sin(Q\Delta) - Q\Delta\cos(Q\Delta))$$

(10)

$$4\pi r_0^2 [g(r_0) - g_0] = \frac{2}{\pi}\int_0^{Q_{max}} Qr_0[S(Q) - 1]\sin(Qr_0)\, f_{SB}(Q\Delta) dQ$$

(11)

Figure 1 shows the Lorch and Soper-Barney modification functions. Compared to the Lorch function, the Soper-Barney function has a wider central maximum[1] but then attenuates the data at higher $Q$ far more than the Lorch function. If $\Delta$ is set analogously to the procedure used above with the Lorch function then the wider central maximum causes a given value of $Q_{max}$ to result in a larger value for $\Delta$. Setting $f_{SB}(Q_{max}) = 0$ is essential to eliminate the discontinuity in $S(Q)$. This results in $\Delta = 4.49(1)/Q_{max}$, compared to $\Delta = \pi/Q_{max}$ for the Lorch method. In some cases the Soper-Barney function has been implemented with the width $\Delta$ being r-dependent, though we will not pursue that here.

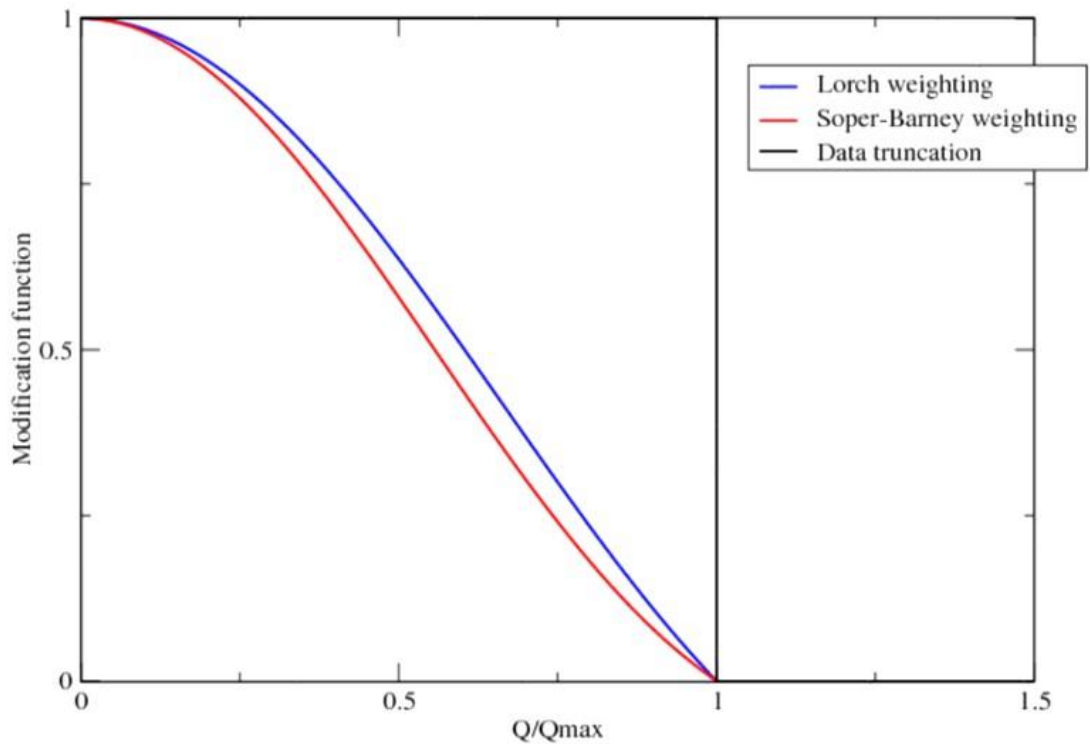

Figure 1. Modification functions which permit the application of equation 4 in the absence of data for $Q > Q_{max}$, plotted as a function of $Q/Q_{max}$. The black line indicates a step function $f_S$ which is discontinuous at $Q_{max}$, a method referred to as "direct FT". The blue line is the Lorch function $f_L$, and the red line is the Soper-Barney function $f_{SB}$. These are discontinuous in the first derivative. All functions continue to infinity with a value of zero.

An alternative to the use of a modification function is to perform the FT (equation 5 or equivalent) to obtain $g(r)$, then perform the inverse FT to return to $S(Q)$, followed by an iterative process to transform back and forth seeking physically reasonable behaviour (in particular, ensuring that $g(r) = 0$ within the atomic radius). This approach was pioneered first by Kaplow et al. [6], then outlined in more detail by Eggert et al. [2]. The adjustable fitting parameters in this process are $g_0$ (representing the density) and two scaling factors specific to their method for background subtraction.

Unfortunately, to perform this analysis it was necessary to smooth the original $S(Q)$ data using a cubic-spline smoothing routine in which the amount of smoothing applied varied as a function of Q. In addition, the optimum values of the density and scaling factors varied substantially according to the value chosen for $Q_{max}$, not converging to a stable value (let alone the correct value for the density) within the range of $Q_{max}$ studied ($6 - 10$ Å$^{-1}$). Due to these difficulties (in particular, the need for $Q$-dependent smoothing) we have not pursued the iterative approach in the present work. Iterative approaches are however available in the software packages Amorpheus [7] and LiquidDiffract [8].

---

[1] To second order in $Q\Delta$, $f_L = 1 - \frac{(Q\Delta)^2}{3!}$ And $f_{SB} = 1 - \frac{(Q\Delta)^2}{10}$

## IV. A Bayesian View on the modification functions

The Lorch method tackled the discontinuity problems (3 and 4 above) by modifying the measured $S(Q)$ by a function which goes to zero at $Q_{max}$. This has the cost that the data has to be adulterated to allieviate the problems of the discontinuity.

Another way to write the exact same transformation is:

$$
\begin{aligned}
4\pi r^2[g(r) &- g_0] \\
&= \frac{2}{\pi} \int_0^\infty \left[ Qr[S(Q)-1]\sin(Qr)f_B(Q/Q_{max}) \right. \\
&\quad + \left. Qr[S_p(Q)-1]\sin(Qr)[1-f_B(Q/Q_{max})]\right] dQ
\end{aligned}
$$

$$(12)$$

With $S_p(Q) = 1$. The second term is introduced to connect the modification approach to a Bayesian framework. In the Bayesian approach to data, one starts with a prior assumption $[S_p(Q)-1]$ for the data, and this is modified by the data. As well as the prior, an essential ingredient for the Bayesian method is the "strength" of the prior.

Taking $f_B$ as the step function and the prior as "zero signal", we can read this equation as placing full weight on the data where it exists, up to $Q_{max}$, then full weight on the prior beyond $Q_{max}$. The Lorch and Soper-Barney modification can be seen as placing some weight on the prior at all Q. e.g. at $Q = Q_{max}/2$, these methods determine $g(r)$ by placing approximately equal weight to the prior as the real data.

Of course, there is no requirement for the prior to be chosen as zero, one might e.g. choose the results of a simulation, or a previous experiment. Whatever prior is chosen, where one has full confidence in the data one can set $f_B = 1$ and the prior will be ignored, if one has no confidence in the data, e.g. because it does not exist, the non-zero prior means that the second term can provide non-zero contribution to the FT from $Q > Q_{max}$.

There are several approaches to the use of the prior: it can be used to combine theory and experiment with weights determined by the researcher, or it can be deployed in the spirit of Lorch/Soper-Barney to remove the discontinuity.

In this second case, one might choose $f_B$ to be the step function, and the prior to be a function which matches the data at $Q_{max}$ and decays gracefully to zero (the form of the prior at $Q < Q_{max}$ is irrelevant, because it will be multiplied by zero). Such an approach retains the advantages of Lorch and Soper-Barney in addressing the $Q_{max}$ cutoff problem, but does not compromise the integrity of the measured data.

## V. The effect of noise in the $S(Q)$ data

Even if data are available up to extremely high $Q$, encompassing all oscillations in $S(Q)$, there remains the problem of noise. From the Debye scattering equation (equation 1), we expect that the physically significant oscillations in $S(Q)$ will decay in amplitude upon $Q$ increase. However, the noise amplitude in real experimental data does not decay on $Q$ increase. In equation 4, $S(Q)$ is multiplied by $Q$ in order to construct a valid FT to invert. The effect of this at high $Q$ is simply to amplify noise. As we see throughout this work, it is common practice to smooth $S(Q)$ data so this problem can be avoided – including implementation of $Q$-dependent smoothing. This, however, raises a separate set of uncertainties: How can one be certain that features of physical significance have not also been lost in the smoothing process?

## VI. Binning requirements for numerical integration to obtain $g(r)$

Regardless of whether we employ the standard FT method, Lorch modification function or Soper-Barney modification function, $g(r_0)$ is obtained by a numerical integration to $Q = Q_{max}$ of a function including $\sin(Qr_0)$. We therefore require the $S(Q)$ data to be binned at sufficiently fine resolution in $Q$ such that a single cycle in the $\sin(Qr_0)$ function covers a large number of $Q$-values in the experimental data. Thus if the binning interval is $\delta Q$ we require:

$$r_0 \delta Q \ll \pi$$

(13)

Thus we can see that the finite binning interval in the $S(Q)$ data can cause the numerical integration procedure to fail when used at large $r_0$.

## VII. Calculation of co-ordination number (CN).

The co-ordination number (CN) can be obtained from diffraction data via $g(r)$ using equation 14 below.

$$\text{CN} = 4\pi\rho \int_{0}^{r_{max}} r^2 g(r) dr$$

(14)

Here, $r_{max}$ is the radius of the minimum following the first and highest peak in $g(r)$. However, as shown in equation 14, this method can only be used if the density $\rho$ is known reasonably accurately.

## Methods

We analysed four different sets of neutron / X-ray diffraction $S(Q)$ data (described later). The numerical integrations to perform the FT and CN calculations were performed using our Octave code, which is available (with documentation) in the supplementary information. When using the Lorch and Soper-Barney modification functions $g(r)$ was obtained from $G(r, \Delta)$ via the approximate relationship in equation 9. In all cases $g(r)$ was calculated for 1000 values of $r$, up to a maximum of 20 Å. Equations 5, 6 and 11 fail if we attempt to calculate $g(r = 0)$ so the minimum value of $r$ was therefore 0.02 Å. Normalization was performed automatically by the code, simply by addition / subtraction to ensure $g(0.02\text{Å}) = 0$ then division by $g(20\text{Å})$ to ensure $g(r) = 1$ in the high $r$ limit. In cases where ripples in $g(r)$ in the low $r$ region caused $g(0.02\text{Å}) > g(20\text{Å})$ normalization and calculation of the CN were not attempted.

The CN was obtained by numerical integration of the first peak in $g(r)$. The integration limits for the calculation of the CN were obtained automatically by the code by locating the maximum value of $g(r)$, then the adjacent minimum values and integrating between these limits. This was chosen instead of integrating from $r = 0$ to reduce the effect of unphysical ripples in $g(r)$ at low $r$.

In all cases (see introduction part IV) the finite binning in the $S(Q)$ data could cause the $g(r)$ obtained at large $r$ to be unreliable. When FT has been performed using the Lorch or Soper-Barney modification function, $g(r)$ may be unreliable at low $r$ (we require $r \gg \Delta$). Allowing ourselves a factor of 10 in these criteria we obtain the conditions given in equation 15 below, which are marked on all graphs of $g(r)$ where they are within the range covered in the plot.

$$r_{max} = \frac{\pi}{10\delta Q}$$

$$r_{min} = \frac{10\pi}{Q_{max}}$$

(15)

The data sets were as follows:

- **Dataset A** is a neutron $S(Q)$ from liquid Ar at 85 K, ambient pressure (a density of 2.13 x $10^{28}$ at./m³) from ref. [9]. The data are collected to 12 Å⁻¹, covering all features in $S(Q)$. The binning is $\delta Q = 0.0294$ Å⁻¹, leading to $r_{max} = 10.7(1)$ Å. The data were smoothed by the authors of ref. [9] using a method communicated to them via private communication, and not described in ref. [9].

- **Dataset B** is a set of synchrotron X-ray diffraction $S(Q)$ data from fluid Ar at 300 K covering eight pressures from 46 MPa to 830 MPa (densities from 9.89 x $10^{27}$ at./m³ to 2.59 x $10^{28}$ at./m³). The binning is $\delta Q = 2.1 \times 10^{-3}$ Å⁻¹, leading to $r_{max} = 149.6(1)$ Å. Since these data have not been published previously we outline here how $S(Q)$ was obtained from the raw data. The raw coherent scattering intensity $I_{raw}(Q)$ was obtained from the experimental data by subtraction of the background signal from either the empty DAC or the DAC containing solid Ar following masking of the solid Ar Bragg peaks. The structure factor $S(Q)$ was obtained by obtaining $I_{raw}(Q)/f(Q)^2$ and normalising (i.e. accounting for the factor of $N'$ in equation 3) to ensure appropriate behaviour in the high-$Q$ limit. We have three comments to make on this process:

1. The $f(Q)$ values are obtained from tabulated data in the International Tables for Crystallography [1]. To obtain $f(Q)$ at the exact values of Q corresponding to our experimental data it is necessary to find an empirical equation that fits the tabulated data, or to do a linear interpolation between the datapoints. We used a linear interpolation. The easier alternative of fitting an empirical equation has some pitfalls. The tabulated $f(Q)$ data in the literature are far more closely spaced at low $Q$, so it would be easy to overfit to these data at the expense of a good fit to the more sparse data at high $Q$. In addition, since $f(Q) \to 0$ at high $Q$, a small error in $f(Q)$ in absolute terms will be large as a proportion of $f(Q)$ and lead to noise in $I_{raw}(Q)$ being amplified by an arbitrary amount.

2. Since the signal-to-noise ratio becomes very poor due to the decreasing $f(Q)$ by the second peak in $I_{raw}(Q)$ it is challenging to accurately normalise $S(Q)$ in the high Q limit. We normalized by measuring the peak scattering intensity from the second peak in $I_{raw}(Q)$ and the minimum scattering intensity from the trough following this peak and normalizing to make the average of these values equal to 1. Clearly, this normalization choice is arbitrary, however in the absence of S(Q) data encompassing all oscillations in S(Q) there is no possible normalization method that does not involve some arbitrary choice.

3. At the lowest values of Q at which data are present, $I_{raw}(Q) < 0$ in most cases, due to difficulties with the background subtraction procedure. A small constant $(1 - 5\%$ of the peak value of $S(Q))$ was therefore added to set $S(Q) \geq 0$ for all $Q$.

Further information on dataset B, including figures, and experimental details, are given in the supplementary information.

**- *Dataset C*** is a set of neutron diffraction $S(Q)$ data from supercritical fluid Kr [10] covering 17 pressures up to 20 MPa and densities up to 6.2 x 10²⁷ at/m³ at 300 K. The data are collected to $Q = 4$ Å⁻¹. We have analysed selected data from this set using Empirical Potential Structure Refinement (EPSR) [12][13] and published this elsewhere [11]. The binning is $\delta Q = 0.05$ Å⁻¹, leading to $r_{max} = 6.3(1)$ Å. The data were smoothed by visual observation by the original authors prior to presentation in tabulated form in ref. [7].

**- *Dataset D*** is a set of neutron diffraction $S(Q)$ data from supercritical fluid Kr [11] collected up to significantly higher pressure and density at 310 K (7 pressures ranging from 40 MPa up to 200 MPa, 1.62 x 10²⁸ at./m³). We have published an EPSR analysis of all these data elsewhere [11]. The binning is $\delta Q = 0.05$ Å⁻¹, leading to $r_{max} = 6.3(1)$ Å. In contrast to dataset C, these data have not been smoothed.

In each case the pressures have been measured experimentally and the densities calculated from the pressures using the relevant fundamental equation of state via NIST REFPROP [14] (ref. [15] for Ar and ref. [16] for Kr). In the case of dataset C, the densities given in the source (ref. [10]) were calculated from the experimentally measured pressures using an earlier equation of state [17]. We have calculated the measured pressures from the densities using ref. [17] then recalculated the densities using the fundamental EOS [16]. This has resulted in a change of ca. 10% in the higher calculated densities.

The pressure-temperature phase diagrams of Ar (from ref. [18]) and Kr (compiled for this work using the methodology presented in ref. [18]) are given in the supplementary information, with the P,T points marked at which the datasets in the present study were collected.

**Results**

**I. Dataset A**

To begin, we performed a standard FT of the entirety of dataset A. Figure 2 shows the $g(r)$ obtained from this FT, with the original $S(Q)$ shown in the inset. The FT result is in agreement with the $g(r)$ shown in ref. [9] (the source of dataset A). The $g(r)$ obtained is also in agreement with that obtained using real space structure determination implemented in the Dissolve package [19] (see supplementary information), and leads to a CN of 11.9. This is clearly a reasonable value for dataset A, however it is worth noting that even with this very good quality data there are unphysical oscillations present in $g(r)$ for $r < 3$ Å, where $g(r) = 0$ is expected. On one level this does not matter (we do not need a FT to tell us that atoms are not allowed to overlap) but it does introduce an error in the normalization of $g(r)$, and hence CN calculation.

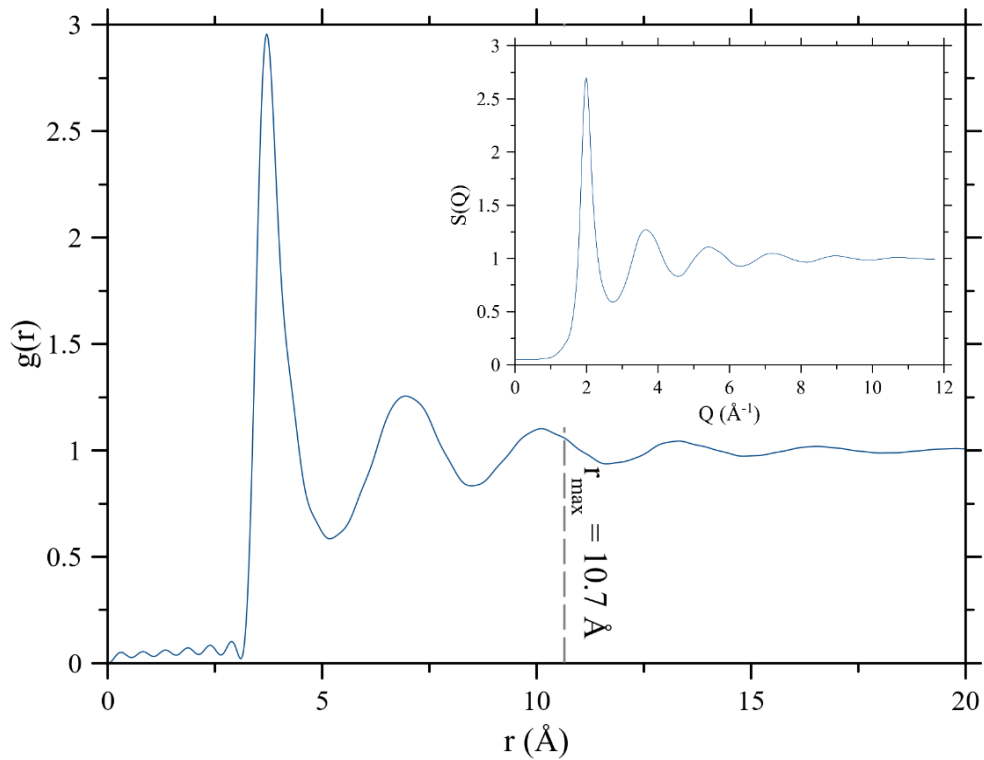

Figure 2. Normalized $g(r)$ for liquid Argon at 85 K (dataset A) obtained by direct FT (i.e. assuming that $[S(Q > Q_{max}) - 1] = 0$) of the full neutron $S(Q)$ (inset).

Attempting the standard FT using cutoffs at successively lower $Q$ leads to the ripples at $r < 3$ Å becoming larger, until at a $Q$ cutoff of 8 Å$^{-1}$ the obtained $g(r)$ is no longer physically realistic, due to the low-r ripples preventing accurate normalization. Figure 3 shows the $g(r)$ obtained with the standard FT, Lorch and Soper-Barney modifications with Dataset A at 8 Å$^{-1}$ Q-cutoff. The Lorch and Soper-Barney modification functions both give reasonable results at cutoffs for which the standard FT fails. The functions give very similar results, the only significant difference being that the principal maximum in $g(r)$ is slightly wider for the Soper-Barney modification function due to the 8 Å$^{-1}$ Q-cutoff resulting in a slightly larger value for Δ.

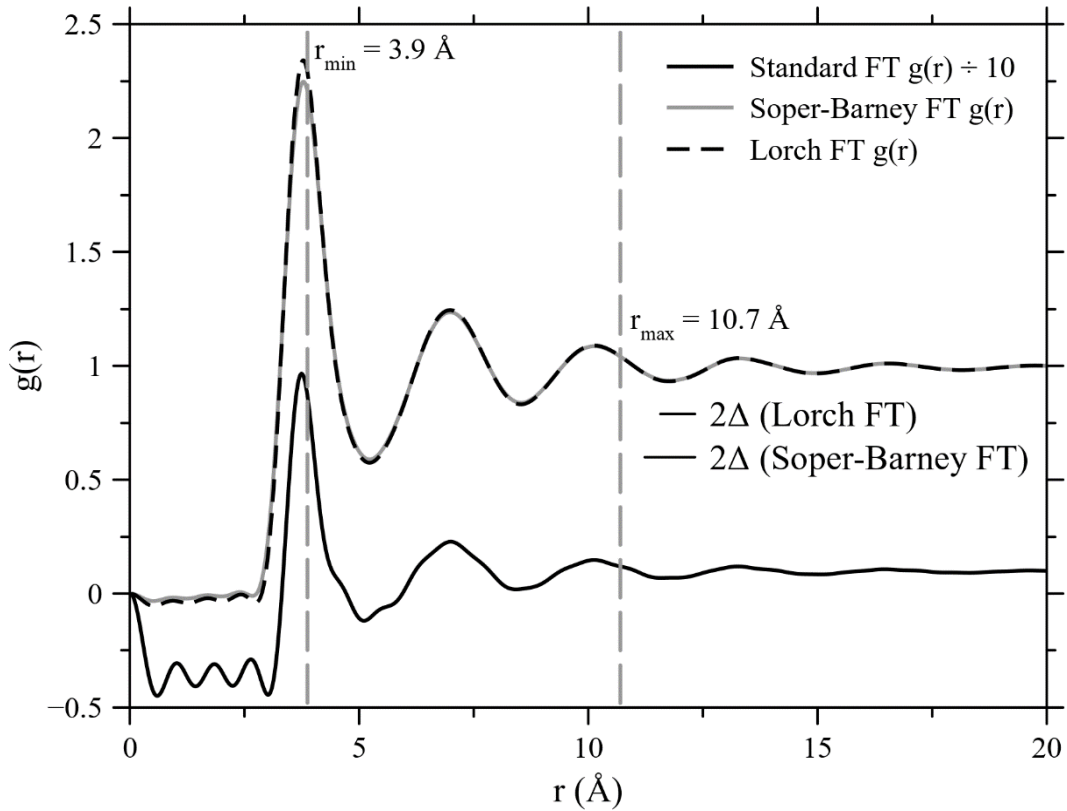

Figure 3. Normalized $g(r)$ for liquid Argon at 85 K (dataset A) obtained from the neutron $S(Q)$ truncated at 8 Å$^{-1}$. FT with abrupt truncation Lorch and Soper-Barney modification are shown. The standard FT $g(r)$ is ÷ 10 for clarity.

Our final investigation with this dataset is to study how low the $Q_{max}$ cutoff can be made with each modification function whilst still obtaining a physically reasonable $g(r)$. Using both the Lorch and Soper-Barney modification functions, even a cutoff as low as 3 Å$^{-1}$ does not result in ripples at low r as bad as those shown for the standard FT in figure 3. A more rigorous test is whether the CN calculated remains reasonable. Figure 4 shows the CNs calculated using all three methods for $Q_{max}$ cutoffs all the way from 3 Å$^{-1}$ to 12 Å$^{-1}$. At 12 Å$^{-1}$ (as previously discussed) all ripples in $S(Q)$ are included in the data and the pure FT gives a $g(r)$ in agreement with that resulting from EPSR analysis of the data. This leads to a CN of 11.9 - as would be expected for a liquid near the triple point. So it is reasonable to take this as the correct value. At 12 Å$^{-1}$ cutoff both modification functions produce a CN slightly larger than the maximum physically realistic value of 12, however as the cutoff is reduced to 5 Å$^{-1}$ the $g(r)$ at least remains stable at this value around 2.5% too large when the FT is done with a modification function. In contrast the CN oscillates wildly when the FT is performed using the standard method.

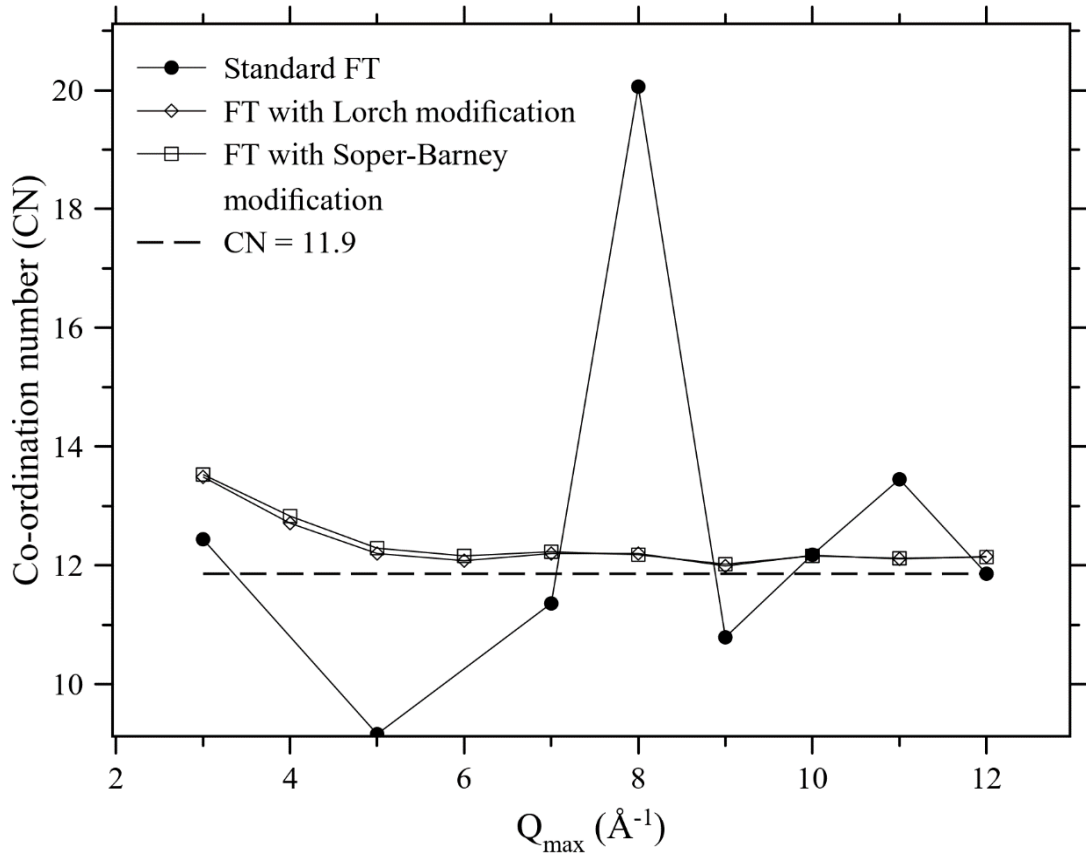

Figure 4.  CN of liquid Ar at 85 K (dataset A) obtained by standard FT of the neutron diffraction data, and FT utilizing the Lorch and Soper-Barney modification functions for a variety of $Q_{max}$ cutoffs.  The $g(r)$ functions produced by the standard FT for 4 Å⁻¹ and 5 Å⁻¹ cutoff could not be normalized so calculation of the CN was not attempted.

## II. Dataset B

Remaining with fluid Ar, dataset B is a set of synchrotron X-ray data (Unlike datasets A, C and D which are all neutron data) collected at 9 pressures up to 830 MPa.  Due to the rapid decrease in scattering intensity upon $Q$ increase (caused primarily by the decrease in the atomic form factor $f(Q)$), the signal-to-noise ratio beyond the first peak in $I_{raw}(Q)$ is very poor.  Only the highest pressure S(Q) produced a g(r) that could be normalized using the method outlined earlier.  Even in this case, to achieve a g(r) that could be normalized it was necessary to adopt the Lorch or Soper-Barney modification function and to use a $Q_{max}$ cutoff that only included the first peak in S(Q).

Figure 5 (a) shows the normalized $S(Q)$ functions at 830 MPa and 444 MPa, illustrating the difficulties caused by the need to normalize on the basis of noisy data at high $Q$.  The first peak in $S(Q)$ is more intense at the lower pressure.

To obtain any reasonable $g(r)$ function it was necessary to use the $S(Q)$ data only up to a $Q_{max}$ at the minimum after the first peak in $S(Q)$ (i.e. ca. $2.5 - 3$ Å⁻¹).  For this reason, analysis could only be performed using the Lorch and Soper-Barney modification functions.  Even with this methodology, only the highest pressure yielded a $g(r)$ that could be normalized.  The obtained CN varied massively depending on which modification function was used: 25.4 with the Lorch modification function and

19.7 with the Soper-Barney modification function. Both of these are far larger than is physically realistic.

Figure 5 (b) shows the normalized g(r) obtained using FT with the Soper-Barney modification function at 830 MPa. Although the normalization procedure is mathematically valid, it is clear that due to the size of the ripple at low r and the existence of negative g(r) in some regions the resulting function is not physically realistic.

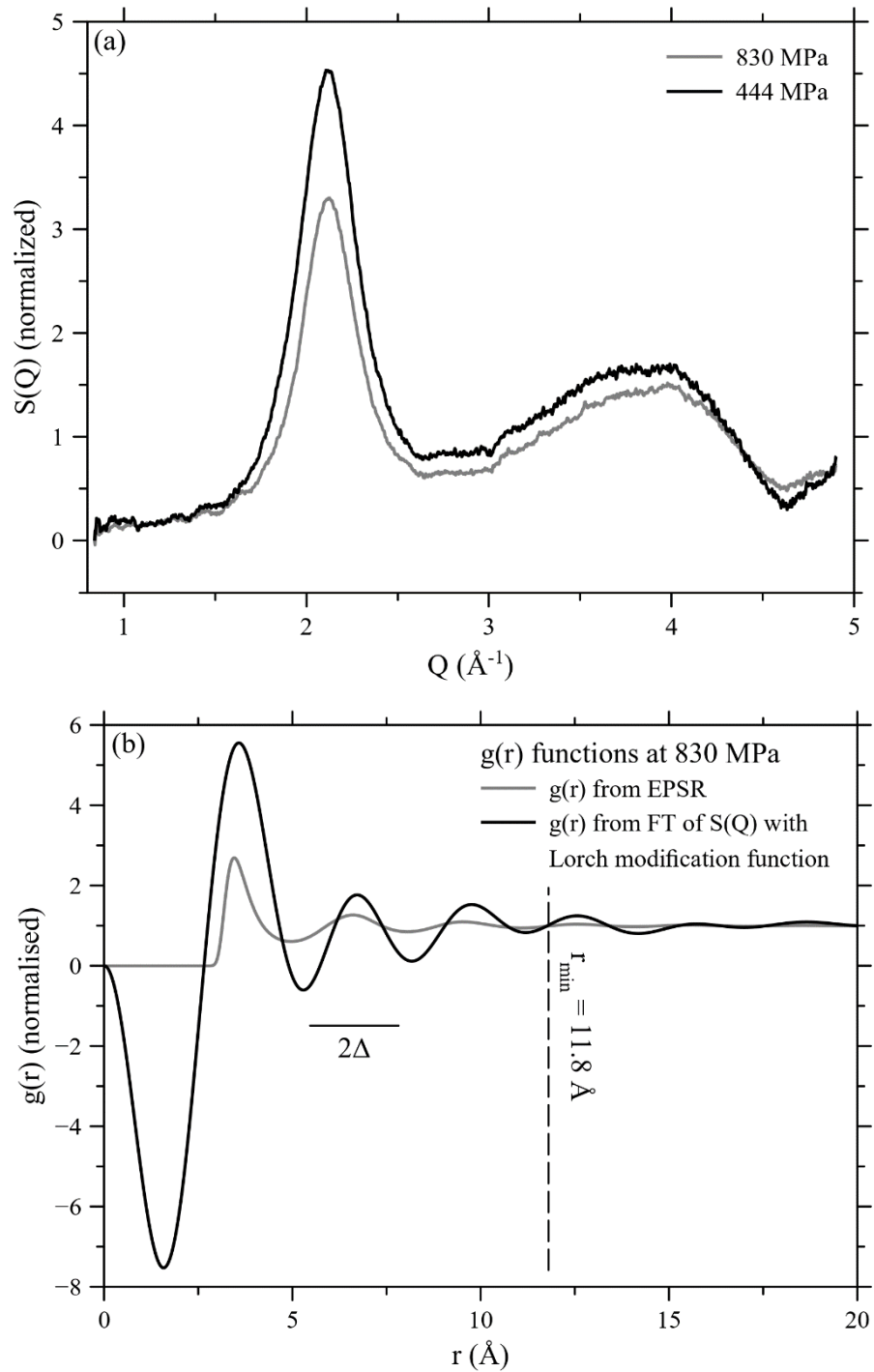

Figure 5. (a) Normalized $S(Q)$ from dataset B at 830 MPa and 444 MPa. (b) $g(r)$ functions at 830 MPa produced from FT of the $S(Q)$ with Lorch modification (cutoff of 2.66 Å⁻¹) and from EPSR using $I_{raw}(Q)$.

We have also performed a full EPSR refinement of the datasets at 830 MPa and 444 MPa, with densities of 43049 Mol./m³ and 37341 Mol./m³. Due to the aforementioned issues with the normalization of the $S(Q)$, we have decided to entirely bypass this and provide to EPSR directly the $I_{raw}(Q)$ as measured. In principle, any lack of correction applied to the $I_{raw}(Q)$ should come out of EPSR as a meaningful trend in the residuals. We employed a 5000 Ar atom simulation box and used the OPLS-Noble Gases forcefield as the reference potential (the same potential parameters as included in Dissolve). Once the box equilibrated, empirical potential (EP) refinement was allowed until a fit that was deemed satisfactory was achieved. At this point, the EP was frozen and prevented from further changing, and the structure was sampled over 5000 accumulations. We achieved an excellent quality of fit to the data, yielding a physically reasonable $g(r)$ and an associated CN of 11.42(15) at 444 MPa. This is below 12, not exceeding the maximum justifiable coordination and is what we would expect on the liquid-like side of the Frenkel line. As expected, the residuals from the fitting process show a clear functional $Q$-dependence, highlighting the lack of appropriate corrections that should have been performed to obtain $S(Q)$ from $I_{raw}(Q)$. An example $g(r)$ resulting from the EPSR is shown in figure 5 (b) and an example S(Q) (with residuals) is shown in figure 6.

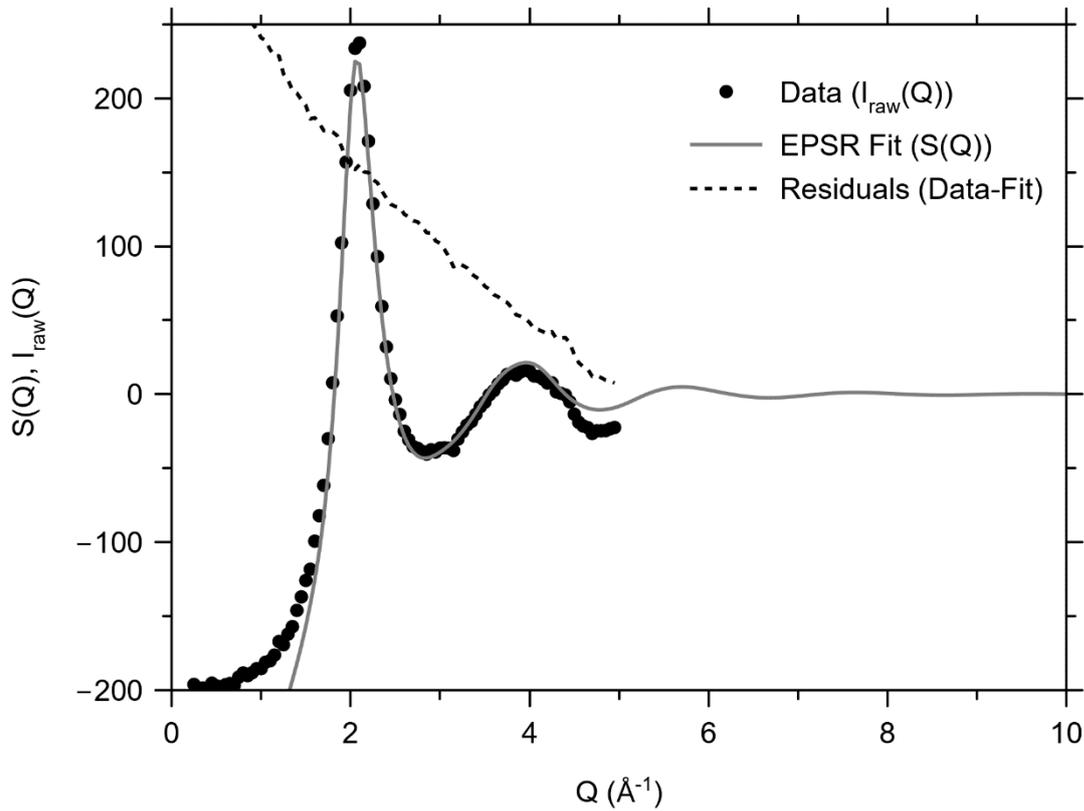

Figure 6. The experimental $I_{raw}(Q)$ data and the fitted $S(Q)$ from EPSR at 444 MPa, demonstrating the excellent quality of fit achieved by EPSR when given the measured $I_{raw}(Q)$, and residuals highlighting the lack of corrections.

### III. Dataset C

Datasets C and D are both from fluid Kr. Dataset C is collected to a relatively low value of $Q$ (4 Å⁻¹), at which significant oscillations in $S(Q)$ are still present. It is therefore not surprising that the standard FT produces a physically unrealistic result, compared to performing the transform using the Lorch method (figure 7 shows the transforms for the highest density point in this dataset, and the $S(Q)$ data). The standard FT produces massive oscillations in the low $r$ limit, and also smaller oscillations in the high $r$ limit which are not physically realistic given the gas-like density (less than half the density of the Frenkel line in Kr at 300 K). Due to the fact that the $S(Q)$ data extend only to 4 Å$^{-1}$, combined with the wide binning interval, the low and high-r regions in which $g(r)$ may be unreliable due to binning and $Q$-cutoff issues respectively overlap for these data.

Figure 8 shows the CNs obtained by Fourier transform of Dataset C at densities from $2.8 - 6.1 \times 10^{27}$ at/m$^3$. This corresponds to pressures up to ca. 20 MPa, not reaching the liquid-like side of the Widom lines, let alone the Frenkel line. We would therefore expect the CN to vary in a smooth and monotonic manner throughout. This is the case when the Lorch or Soper-Barney modification functions are employed. Data are also available in ref. [10] at four lower densities, however none of the FT procedures utilized here produced physically realistic $g(r)$ functions in these cases. The standard FT resulted in a $g(r)$ that could not be normalized, whilst the Lorch and Soper-Barney modification functions produced a $g(r)$ with no clear minimum following the first peak. This resulted in subsequent peaks also being included in the integration to obtain the CN and an unrealistically high value being obtained.

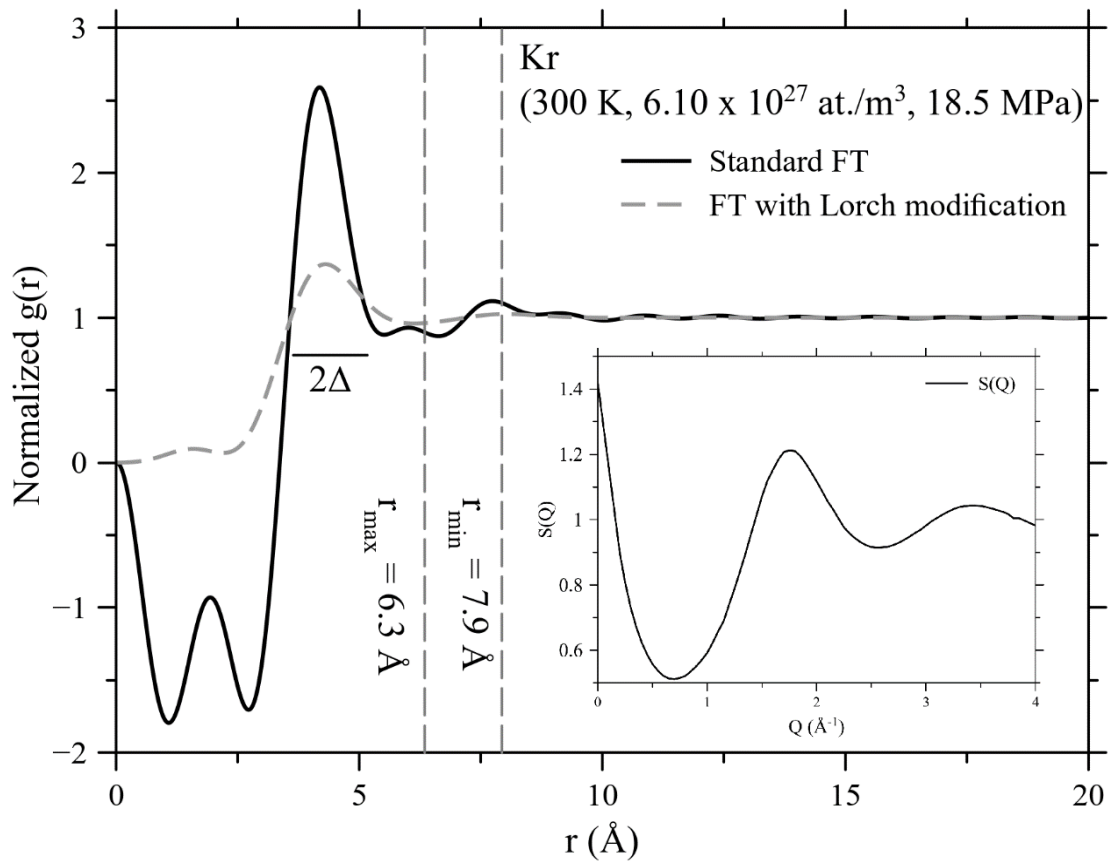

Figure 7. Normalized $g(r)$ obtained from direct FT and FT with Lorch modification of neutron $S(Q)$ from supercritical fluid Kr at 300 K (6.10 x 10$^{27}$ at/m$^3$ density) [10]. Inset: $S(Q)$ data from ref. [10].

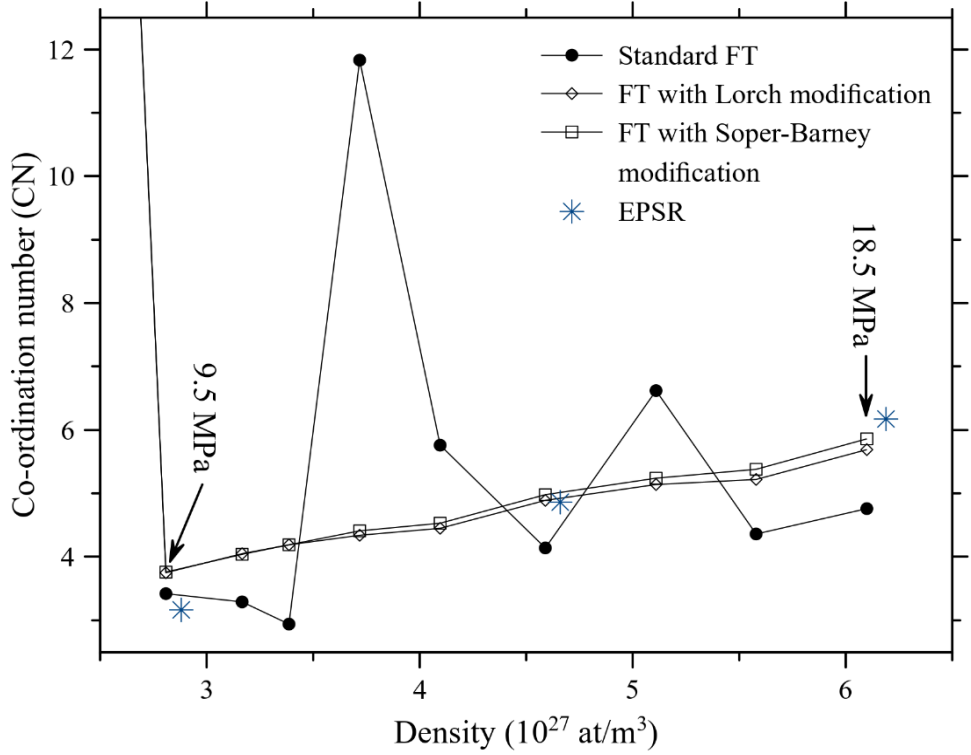

Figure 8. CNs obtained by Fourier transform (present work) and EPSR (ref. [11]) of dataset C (neutron diffraction data from Kr at 300 K to 4 Å⁻¹ from ref. [10]).

We have published an analysis of selected neutron diffraction data from ref. [7] elsewhere [11]. The CNs obtained from EPSR are in reasonable agreement with those obtained by direct FT via both Lorch and Soper-Barney modification functions (figure 8). It is hard to discern the extent to which quantitative agreement exists between the $g(r)$ functions produced by modified FT and EPSR due to the $\pm \Delta$ broadening in $r$ caused by the modification process (figure 7).

### IV. Dataset D

Finally, we present our analysis of dataset D. The large amount of noise in this data prevents meaningful results being obtained by direct Fourier transform, despite the large $Q_{max}$ (20 Å⁻¹). Beginning with a high density datapoint (176 MPa (1.57 x 10²⁸ at./m³) we can see that above 6 Å⁻¹ the data are just noise (figure 9 inset shows the S(Q) data to 15 Å⁻¹). In figure 9 we show $g(r)$ functions obtained with the Soper-Barney modification function for two cutoffs (6 Å⁻¹ and 15 Å⁻¹). The data for 15 Å⁻¹ are ruined by ripples resulting from Fourier transforming noise, whilst the data for 6 Å⁻¹ exhibit a principal peak in $g(r)$ that is somewhat larger than the ripples and leads to a CN of 12.4 (our EPSR analysis of this data published previously indicated a CN of 11.7). The highest density datapoint (200 MPa) can be analysed in a similar manner (CN of 12.5) but for the next lowest pressure (148 MPa) the Fourier transform produces a $g(r)$ that cannot be normalized.

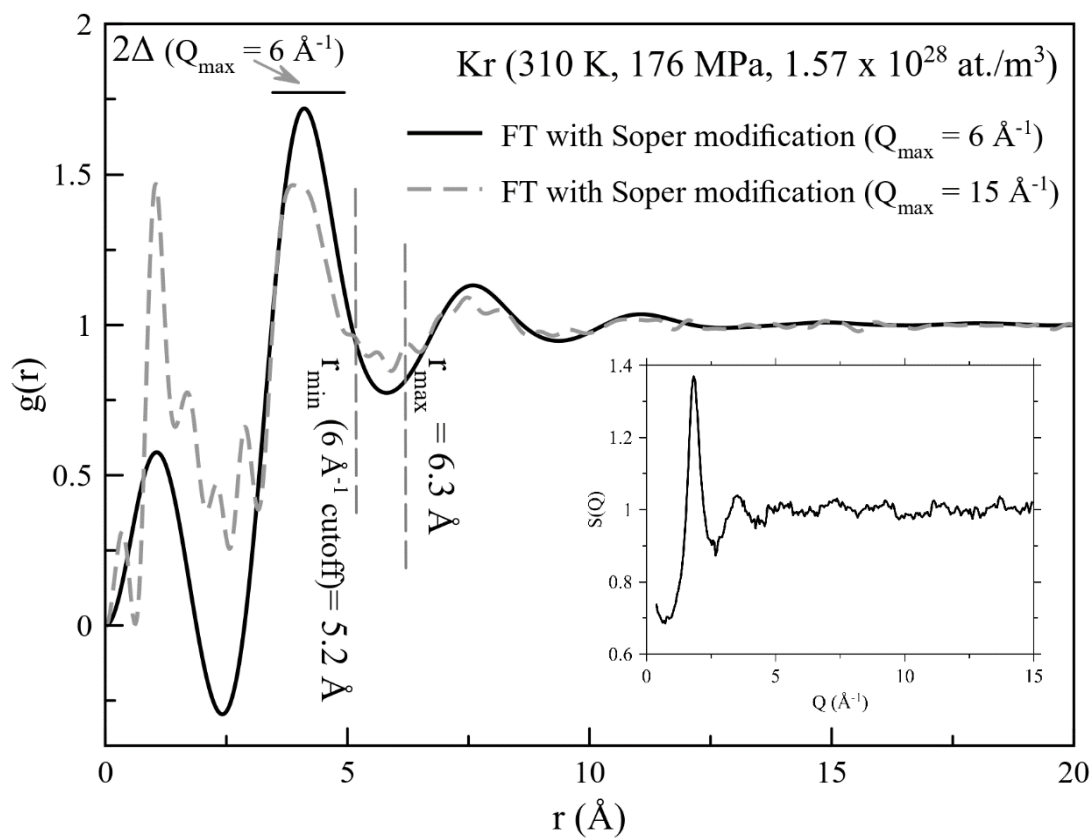

Figure 9. $g(r)$ functions for fluid Kr at 310 K, 176 MPa obtained by FT with Soper modification function with different cutoffs as indicated. Inset: $S(Q)$ data for Kr at 310 K, 176 MPa.

## Conclusions

The results and analysis presented here throw light on some important, but hitherto neglected, aspects of the analysis of fluid X-ray and neutron diffraction data by FT from reciprocal space to real space. In the analysis of X-ray data, and even (to some extent) neutron data, various arbitrary choices are unavoidably made in the analysis. It is therefore difficult to judge the reliability or otherwise of $g(r)$ functions generated by FT of diffraction data without knowledge of the details of these arbitrary choices: In particular, the methodology used to normalise $S(Q)$ and $g(r)$, and the methodology used to obtain $f(Q)$ at the required Q-values from the tabulated data in the literature. In addition, it is essential to know the resolution in the binning of the experimental data as this determines the validity of the resulting $g(r)$ function at high $r$.

Moving beyond these issues, the analysis of diffraction data by direct FT, or by FT using a modification function, is always subject to some extent to the $Q_{max}$-cutoff problem. We find many cases where the Lorch and Soper-Barney modification functions can produce a non-pathological $g(r)$ whilst direct FT cannot and despite concerns raised in recent years, the Lorch modification function is a mathematically rigorous method to analyse fluid diffraction data.

However, mathematical rigor is no guarantee of physical correctness. We have identified four ways in which the $Q_{max}$-cutoff problem can produce physically unreasonable $g(r)$ functions when diffraction data from disordered materials are analysed by FT, but only one of these (the continuity of

$S(Q)$ at $Q_{max}$) is addressed by the Lorch and Soper-Barney modification functions. Neither method attempts to address the issue of missing data beyond $Q_{max}$, the discontinuity in slope of $S(Q)$. Neither method guarantees that the $g(r)$ corresponds to any possible arrangement of atoms in space, certainly not the correct one. Neither method guarantees positive $g(r)$, nor zero $g(r)$ at unphysically small distances. We found that the $g(r)$ functions produced by these methods are highly sensitive to choice of $Q_{max}$, and strong evidence that the results have not converged for typical values of $Q_{max}$.

The fundamental problem with purely data-driven Fourier methods is that they necessarily require some theoretical assumption about the missing data above $Q_{max}$. All the methods used here assume that $[S(Q > Q_{max}) - 1] = 0$, and seek only to address the discontinuity. These mathematical tricks remove some pathological errors in $g(r)$, but the sensitivity to $Q_{max}$ gives little confidence that the non-pathological $g(r)$ are accurate.

The mathematical elegance of these models should not disguise the fact that they serve to suppress the high-$Q$ data. It can be argued that measurement of $[S(Q) - 1]$ may be less reliable at high-$Q$, however the solution to unreliable data is to downweight its importance in analysis, not to pretend that it is smaller than measured, or even zero, which is what the Lorch and Soper-Barney methods do. Indeed, by "modifying" the data away from the experiment in a way designed to avoid pathologies, rather than based on physics, an inaccuracy (broadening) in $g(r)$ is caused.

The problem of missing or unreliable data can be approached with Bayesian-type methods. These provide a framework to combine experimental and theoretical results with appropriate $Q$-dependent weights. They also indicate a way to remove the $Q_{max}$ discontinuities without modifying the data.

Nevertheless, direct FT methods can address only two of the four constraints on $g(r)$ which must be satisfied. To our knowledge, there is no method to determine whether a given function "$g(r)$" or "$S(Q)$" can correspond to any 3D ensemble of atomic arrangements. Therefore, assuming one wishes the data analysis to produce a $g(r)$ which respects the atomic theory of matter, one needs to use a method beginning from real-space configurations. EPSR [12][13] is an example of this approach. It exploits the fact that for any given interatomic potential, $V(r)$, both $g(r)$ and $S(Q)$ are uniquely defined, and respect the atomic theory of matter. So EPSR seeks to find a $V(r)$ consistent with $S(Q)$, and a $g(r)$ which automatically satisfies all four unphysicality issues. EPSR has elements of the Bayesian approach, with the prior being a potential which is modified by the data to produce a best fit to the measured $S(Q)$ from functions constrained to be physically reasonable.

Comparison to EPSR results for all 4 datasets has shown that EPSR can provide physically realistic results even for very poor quality data for which analysis by FT fails. It appears that, historically, $S(Q)$ data have often been smoothed to allow analysis by FT. Results presented here indicate that EPSR can provide physically realistic results with the unsmoothed data. Therefore, if smoothing of the $S(Q)$ data is necessary to allow analysis by FT then EPSR should be attempted instead.

Finally, whilst the Lorch and Soper-Barney modification functions make some progress towards addressing the $Q_{max}$-cutoff problem, neither address the other cutoff problem: The fact that real $S(Q)$ data do not extend to $Q = 0$. In any real diffraction experiment, $I_{raw}(Q)$ data cannot be collected at $Q \approx 0$ as there is no way to distinguish between X-rays / neutrons that have been transmitted without interacting with the sample, and those scattered at low $Q$. Whilst $\lim_{Q \to 0} S(Q) \neq 0$, the contribution to the value of $[g(r) - g_0]$ from low-$Q$ scattering in equation 4 (and modified FT functions derived from it) *does* vanish due to the presence of the $\sin(Qr)$ term in the integral so the substitution of the zero lower limit in the integral with a small but finite minimum value of $Q$ should be acceptable. However, it is worth noting that analysis methods beginning from real-space configurations and predicting $S(Q)$ in the $Q$-range for which data exists also avoid this potential pitfall.

In future the work presented here can be extended to cover other fluids comprised of spherically symmetric particles such as metallic fluids and $CH_4$, and molecular fluids for which different partial $S(Q)$ data are obtained by isotopic substitution in neutron diffraction.

## Supplementary material

Supplementary material is provided, consisting of: $g(r)$ function obtained by performing EPSR on dataset A, full description of the experimental methods and analysis to obtain $S(Q)$ for dataset B, the Octave / Matlab code used, and documentation for this code.

## Acknowledgements

We would like to acknowledge the provision of beamtime at Diamond Light Source I15 (CY28469-1) and ISIS Pulsed Neutron Source SANDALS instrument (2010332), and acknowledge Prof. Daniel Errandonea for allowing the Ar X-ray experiment to be performed at beamtime CY28469-1. GJA and CP acknowledge funding from the ERC under the Hecate grant. For the purpose of open access, the authors have applied a Creative Commons Attribution (CC BY) licence to any Author Accepted Manuscript version arising from this submission.

## Author declarations

The authors have no conflicts of interest to declare.

Author contributions: JEP designed the study, performed experiments, wrote code, performed data analysis by FT, and wrote the manuscript. CGP designed the study, performed experiments, performed data analysis by EPSR and contributed to the manuscript. BM wrote code and performed data analysis by FT. MAK performed data analysis and prepared figures. GJA designed the study, contributed to the manuscript and prepared figures. CWM prepared figures and SA performed experiments. All authors approved the final version of the manuscript.

## Data availability statement

The data that support the findings of the study are available from the corresponding author upon reasonable request.

## References

[1] A. J. C. Wilson and E. Price, International Tables for Crystallography (Kluwer, Dordrecht/International Union of Crystallography), Vol. C (1999).

[2] J.H. Eggert, G. Weck, P. Loubeyre and M. Mezouar, Phys. Rev. B **65**, 174105 (2002).

[3] B.E. Warren, *X-ray diffraction* (Dover publications, 1990).

[4] E. Lorch. J. Phys. C (solid state physics) **2**, 229 (1969).

[5] A.K. Soper and E.R. Barney, J. Appl. Cryst. **44**, 714 (2011).

[6] R. Kaplow, S.L. Strong and B.L. Averbach, Phys. Rev. **138**, 1336 (1965).

[7] S. Boccato et al., High Press. Res. **42**, 69 (2022).

[8] B.J. Heinen and J.W.E. Drewitt, Phys. Chem. Minerals **49**, 9 (2022).

[9] J.L. Yarnell, M.J. Katz, R.G. Wenzel and S.H. Koenig, Phys. Rev. A **7**, 2130 (1973).

[10] A. Teitsma and P.A. Egelstaff, Phys. Rev. A **21**, 367 (1980).

[11] C.G. Pruteanu, J.S. Loveday, G.J. Ackland and J.E. Proctor, J. Phys. Chem. Lett. **13**, 8284 (2022).

[12] A.K. Soper, Chem. Phys. **202**, 295 (1996).

[13] A.K. Soper, Mol. Sim. **38**, 1171 (2012).

[14] Lemmon, E.; Bell, I. H.; Huber, M.; McLinden, M. NIST Standard Reference Database 23: Reference Fluid Thermodynamic and Transport Properties-REFPROP, Version10.0, Standard Reference Data Program, National Institute of Standards and Technology: Gaithersburg, 2018.

[15] Ch. Tegeler, R. Span and W. Wagner, J. Phys. Chem. Ref. Data **28**, 779 (1999).

[16] E.W. Lemmon and R. Span, J. Chem. Eng. Data **51**, 785 (2006).

[17] N.J. Trappeniers, T. Wassenaar and G.J. Wolkers, Physica (Utrecht) **32**, 1503 (1966).

[18] J.E. Proctor, *The Liquid and Supercritical Fluid States of Matter* (CRC Press, 2020).

[19] T. Youngs, Mol. Phys. **117**, 3464 (2019).

# A comparison of different Fourier transform procedures for analysis of diffraction data from noble gas fluids

## Supplementary information

J.E. Proctor[1], C.G. Pruteanu[2], B. Moss[1], M.A. Kuzovnikov[2], G.J. Ackland[2], C.W. Monk[1] and S. Anzellini[3]

1. Materials and Physics Research Group, School of Science, Engineering and Environment, University of Salford, Manchester M5 4WT, UK

2. SUPA, School of Physics & Astronomy and Centre for Science at Extreme Conditions, the University of Edinburgh, Edinburgh EH9 3FD, UK

3. Diamond Light Source Ltd., Harwell Science and Innovation Campus, Diamond House, Didcot OX11 0DE, UK

## 1. Dataset A real-space structure determination

We performed real-space structure determination on Dataset A using the Dissolve software package [19], producing a $g(r)$ in agreement with that obtained from standard FT of the dataset for the full range in $Q$-values. The simulation employed a 5000 atom simulation box and the OPLS-Noble Gases forcefield potential. It was not necessary in this case to refine the potential. The structure was sampled over 5000 accumulations. Figure S1 shows both $g(r)$ functions.

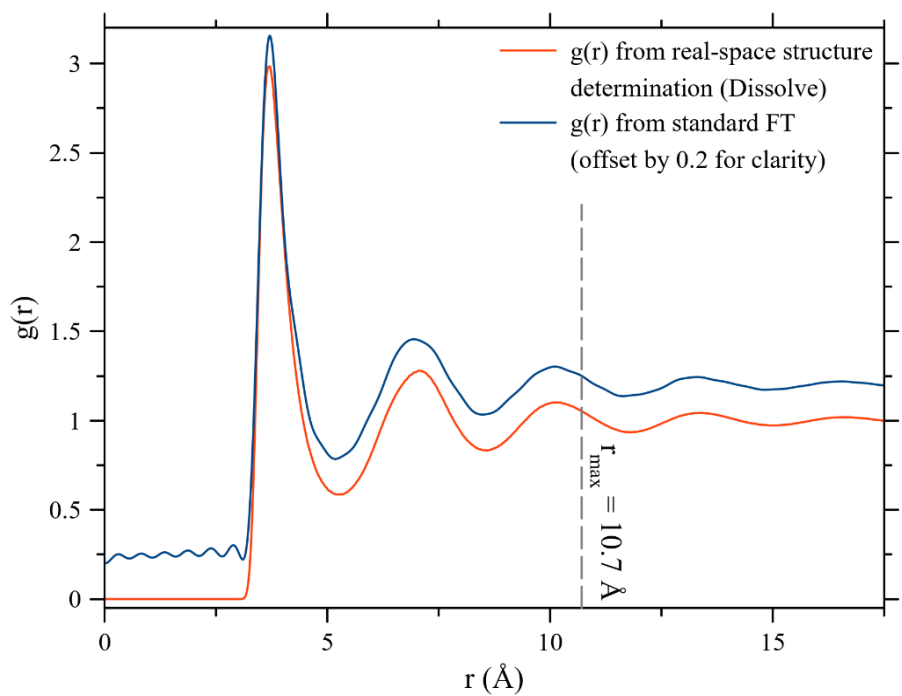

Figure S1. The $g(r)$ functions for Dataset A obtained via standard Fourier transform using the full range of $Q$-values for $S(Q)$, and via real-space structure determination using the Dissolve package.

## 2. Dataset B experimental methods

Pressure was applied using a custom-constructed piston-cylinder diamond anvil cell (DAC) equipped with 600 μm diameter culets and a $2\theta = 20°$ opening on the cylinder (downstream) side. This opening corresponds to $Q = 5.15$ Å$^{-1}$. Data collected at higher $Q$ than this were not utilized in our analysis, therefore it was not necessary to collect / compute a transmission function for the DAC seats as was done in ref. [2]. An indented stainless steel gasket was utilized, in which a hole was prepared using a custom-constructed spark eroder device. Argon (BOC zero grade, 99.999%) was loaded cryogenically by placing the DAC in a small chamber into which Argon was pumped after purging the chamber of air. The chamber was placed in a bath of liquid nitrogen in order to condense the Argon, and the screws were turned to close the DAC whilst it was completely immersed in liquid Argon.

Synchrotron X-ray diffraction data were collected at Diamond Light Source beamline I15 using a 29.3 keV X-ray beam. The beam was focussed to 9 μm x 6 μm (FWHM) and a Pilatus CdTe 2M detector was used. The sample-to-detector distance (approx. 424 mm) and beam energy were calibrated using a CeO$_2$ standard. Typical data acquisition time was 30 s. Azimuthal integration was performed using Dioptas v0.5.5 software. Detector artefacts, Bragg peaks from diamond anvils and shadows from beamstop and other auxiliary equipment were manually masked, and all detector images were processed with the same mask to enable a subtraction of individual patterns. "The X-ray beam was aligned to the centre of the sample chamber before collecting each pattern".

Pressure was measured using the Ruby photoluminescence method. Due to the need to collect many closely spaced datapoints, pressure was measured before and after each X-ray diffraction pattern was collected and data were rejected if the change in pressure during data collection exceeded 30 MPa.

The experiment on I15 in which these data were collected was setup at short notice following an altnernate experiment being aborte for technical reasons. As a result, the I15 beamline was not setup with the optimum X-ray wavelength, geometry and detector for this experiment.

## 3. Dataset B normalization

In pressure points comprising dataset B $I_{raw}(Q)$, the raw coherent scattering intensity from the sample, was obtained from the total raw scattering intensity by subtraction of a background originating from either the empty DAC or the DAC containing Ar in the solid state. Figure S2 below shows examples of both these calculations.

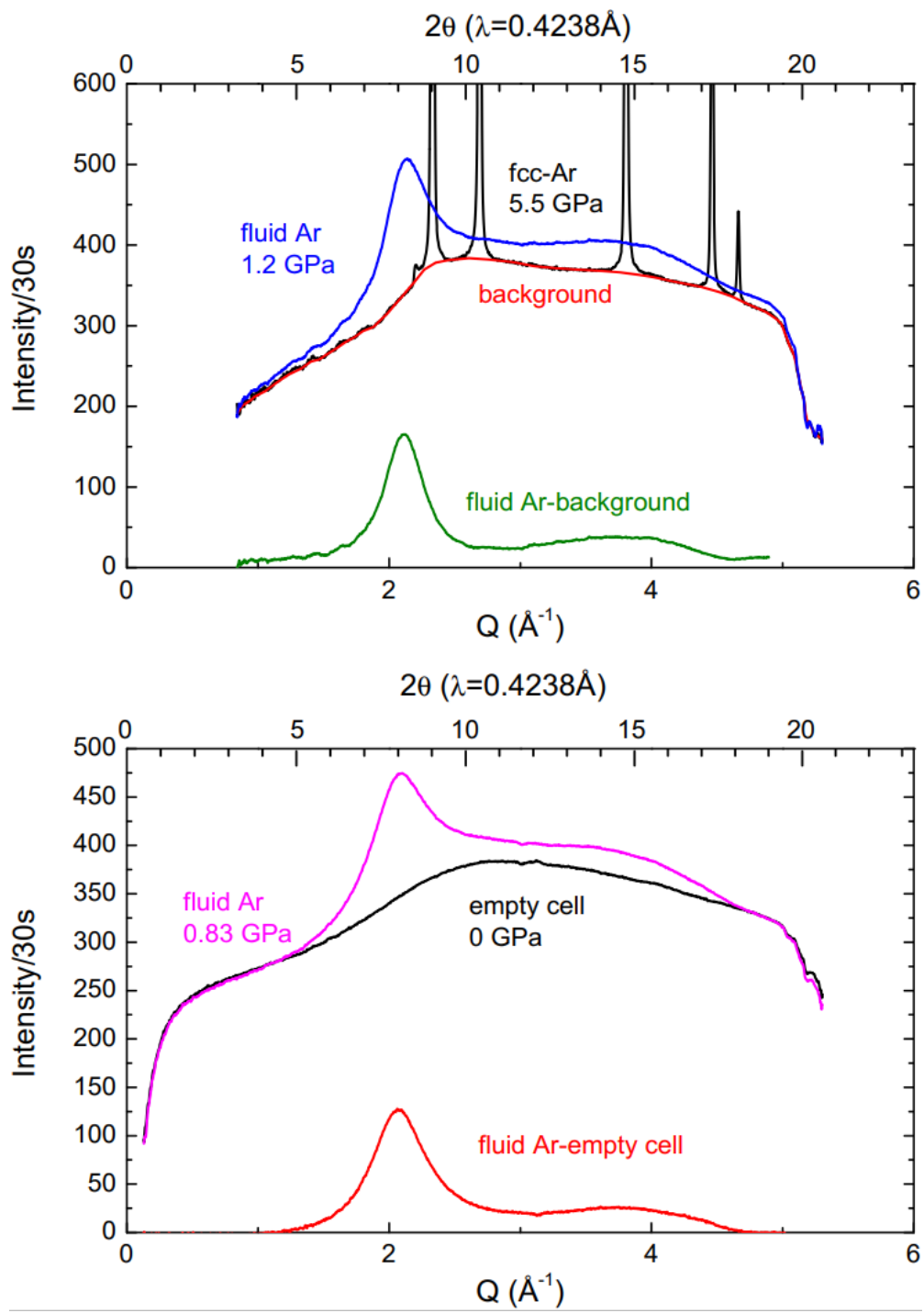

Figure S2. Examples of background subtraction on dataset B to obtain $I_{raw}(Q)$ by subtraction of the background signal from Ar confined in the DAC in the solid state (upper) and from the empty DAC (lower).

Dataset B was normalized according to equation 3, reproduced below:

$$S(Q) = \frac{I_{raw}(Q)}{N' f(Q)^2}$$

The parameter $N'$ is a fitting parameter to ensure that $S(Q) \rightarrow 1$ in the high $Q$ limit, whilst $f(Q)$ is the atomic form factor. The most commonly used set of values are those given in the International Tables for Crystallography [1]. We used the linear interpolations between the values given in the

International Tables for Crystallography shown in figure S3. In practice, performing background subtraction that produced reasonable outcomes at all $Q$ without obtaining negative $I_{raw}(Q)$ for some low values of $Q$ was challenging. The data were therefore shifted by a small constant (less than 5% of the peak value) to ensure that the lowest value of $S(Q)$ was at least zero.

Figure S4 shows, for example data at 830 MPa, the stages in the normalization process beginning with $I_{raw}(Q)$. Firstly, $I_{raw}(Q)$ is divided by $f(Q)^2$, after which it is normalized to 1 in the high-Q limit. Since the value of with $I_{raw}(Q)$ is still oscillating at $Q = 5$ Å⁻¹, this was done by normalizing such that the average value of the second peak (at $Q \approx 4$ Å⁻¹), and the trough following it, was 1.

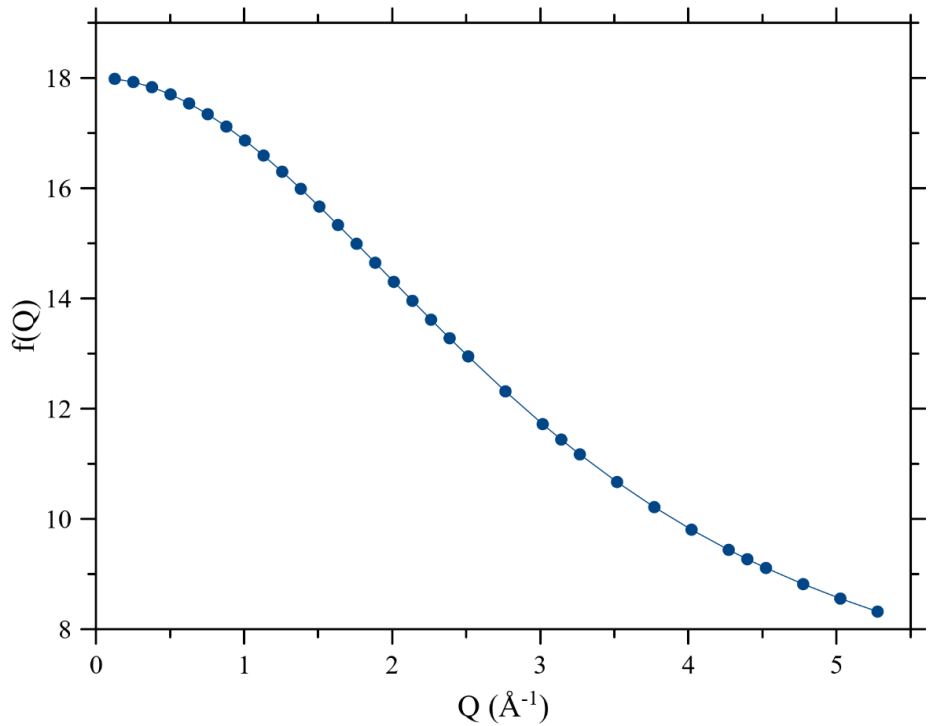

Figure S3. Tabulated f(Q) data from the Intl. Tables for Crystallography [1] (points) and our interpolation (line).

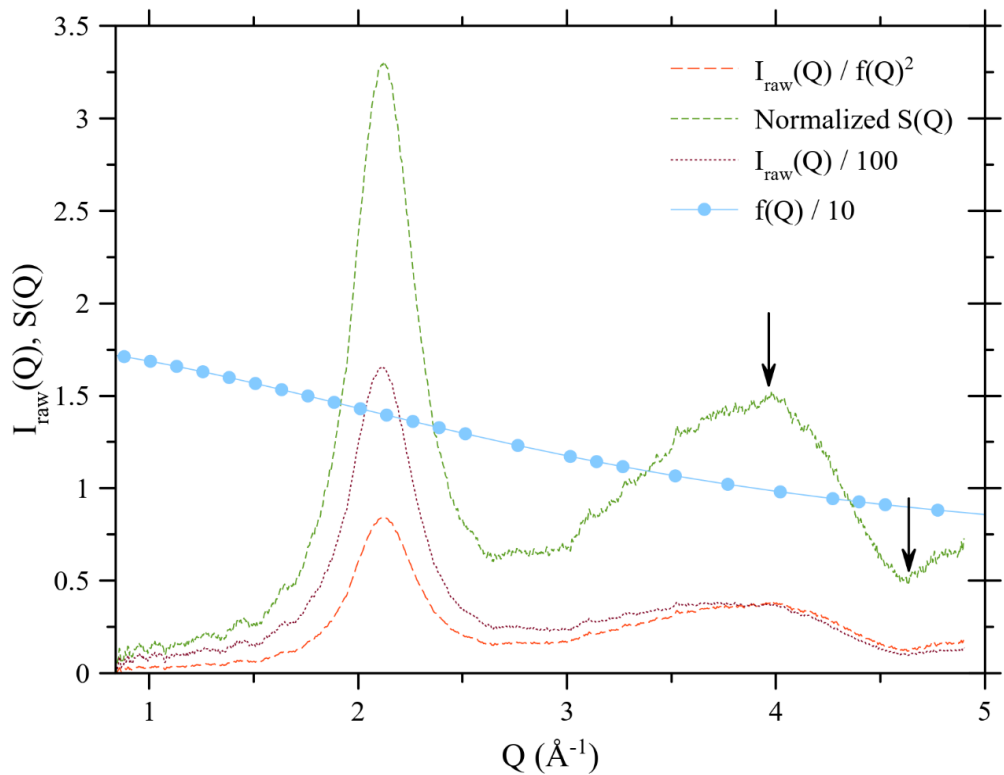

Figure S4. Stages in the normalization process to obtain $S(Q)$ from $I_{raw}(Q)$ at 830 MPa. The division by 100 applied to $I_{raw}(Q)$ and ($I_{raw}(Q)/f(Q)^2$) was performed solely to enable presentation on this figure alongside $f(Q)$ and $S(Q)$, and was not part of the normalization process. The arrows indicate the points selected in the last stage of the normalization process: Ensuring that the average value of S(Q) between these points is 1.

## 4. Phase diagrams

Figure S5 (below) shows the phase diagrams of fluid Ar (from ref. [18]) and Kr (compiled for this work using the methodology presented in ref. [18]), with the P,T points marked at which the different datasets in the present work were collected.

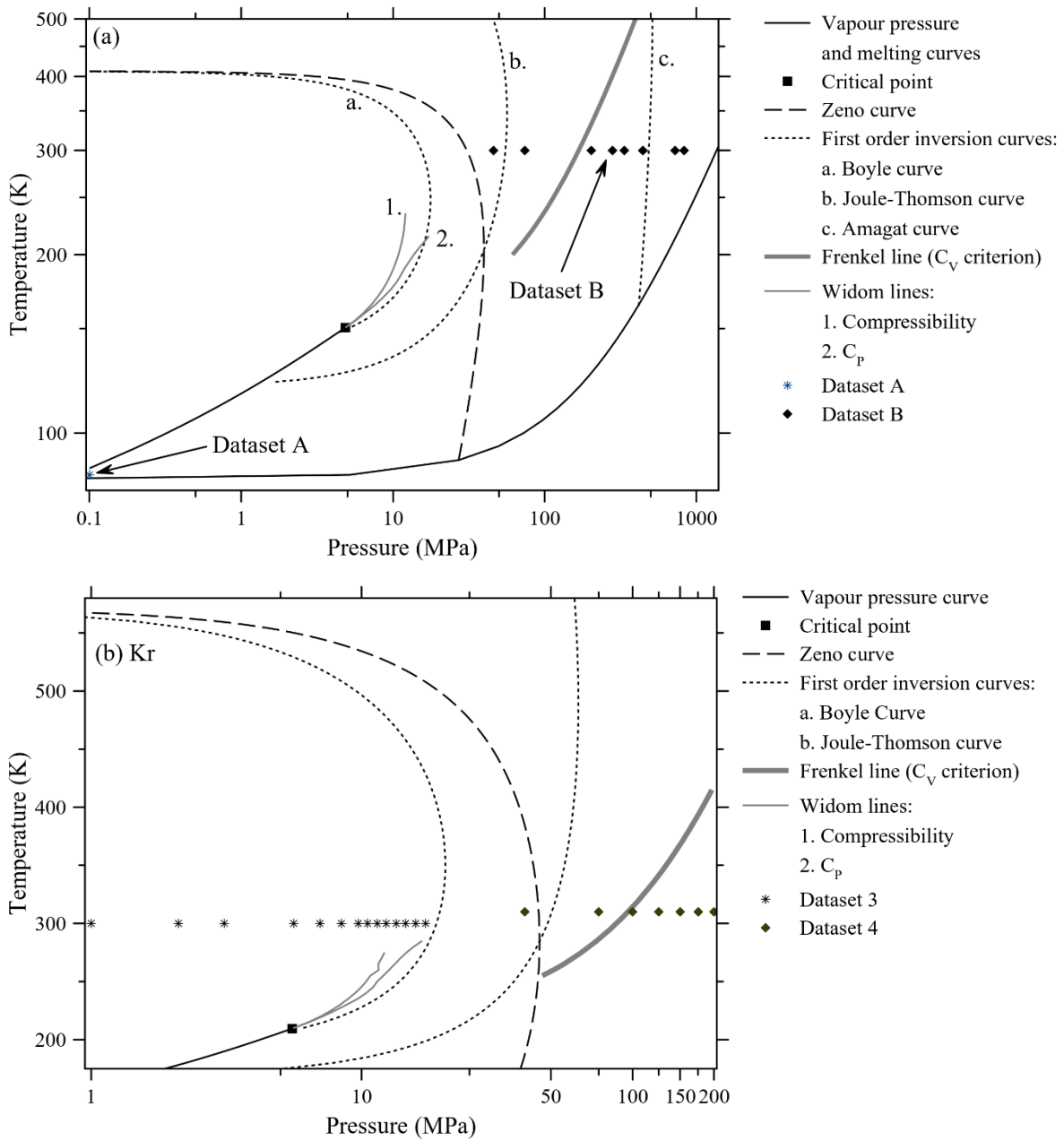

Figure S5. (a) Phase diagram of Ar (from ref. [18] with the P,T points for datasets A and B added) and panel (b) Phase diagram of Kr (compiled using the methodology of ref. [18] with the P,T points for datasets C and D added).

## 5. Code documentation

The Fourier transform (FT) results described in the text were obtained using our own code as documented below. We used Octave (an open-source version of Matlab) throughout. There are three separate scripts, one for the standard Fourier transform and one for each modification function. The code calculates a FT using the full range of the input data (all $Q$ values) and a FT using $Q$ values only up to a certain lower cutoff. The scripts for the modification functions are based closely on the script for the standard Fourier transform so we will document this first. All three scripts are available to download with this supplementary information.

**5.1 Standard Fourier transform**

The $S(Q)$ input data should be provided in tab separated variable format where the first column is filled with the $Q$-values (in Å⁻¹) and the second column with the $S(Q)$ values. All text such as column headings must be deleted from the input data file by the user prior to running the script. The data should be stored in the same directory as the script, in which case to load the data it is simply necessary to write the file name including extension in line 4 and the file name without the extension in line 5.

The following additional data need to be entered by the user:

- Line 6, the number density of particles p in at./m³.
- Line 7 Qmax, the row number in the $S(Q)$ input data file at which to stop for the FT using the cutoff at lower Q.
- Line 9 n, the number of (equally spaced) values of r at which to calculate $g(r)$.
- Line 10, r_max, the maximum value of r (in metres) to which to calculate $g(r)$.

The remaining variables are as follows:

| Name | Purpose |
|---|---|
| Q_values | Array of $Q$ values (read from file in Å⁻¹, converted immediately to m⁻¹ when array is filled). |
| S_Q_values | Array of normalized $S(Q)$ values read from file. |
| r_values | The array of n values of r (in metres) for which $g(r)$ is to be calculated, starting at r_spacing (since equation 4 in the main text fails at $r = 0$) and ending at r_max. |
| r_spacing | the spacing between r values. |
| Q_values_cutoff | Equivalent of Q_values but containing data only up to Q_values(Qmax). |
| S_Q_values_cutoff | Equivalent of S_Q_values but containing data only up to S_Q_values(Qmax). |
| raw_g_r | Array of all (unnormalized) $[g(r) - g_0]$ values for the FT of the full dataset. |
| raw_g_r_cutoff | Equivalent to raw_g_r but for the FT of the data up to Q_values(Qmax). |
| norm_g_r | Normalized g(r) for full FT. |
| norm_g_r_cutoff | Ditto but for cutoff FT (same number of r values in both cases). |
| Q_value | Value of Q at the cutoff point. |
| CN | Calculated co-ordination number (CN) with full range $g(r)$. |
| CN_cutoff | Calculated co-ordination number (CN) with $g(r)$ obtained with $S(Q)$ cutoff at lower Q-value. |

Table S1. Names of principal non-user-defined variables in the code for the standard Fourier transform.

The code obtains the normalized $g(r)$ and the co-ordination number (CN) for the full set of $S(Q)$ data and the data cutoff at the lower value of $Q$ in the following steps. Firstly, the equation below (equation 4 from the main text rearranged and with upper and lower limits inserted in the integral) is used to calculate $[g(r) - g_0]$.

$$g(r) - g_0 = \frac{1}{2\pi^2 r^2} \int_{Q_{min}}^{Q_{max}} Qr[S(Q) - 1]sin(Qr)dQ$$

This is performed for each value of $r$ using the trapz numerical integration function. The parameter $Q_{min}$ is simply the minimum value of $Q$ for which $S(Q)$ experimental data is provided.

The same procedure is used to normalize the $g(r)$ functions for both the full and reduced range in $Q$-values. A constant is added to the value of $g(r)$ at all $r$ to ensure that $g(r_{min}) = 0$ (essentially just removing the arbitrary constant $g_0$). Then the $g(r)$ data at all $r$ are divided by $g(r_{max})$. Clearly, this procedure will not work if the ripples in $g(r)$ at low $r$ result in $g(r_{max}) < g(r_{min})$ so first the code checks for this and displays an error message if this is the case. Clearly it would be possible to contrive other, more complex, procedures for normalizing $g(r)$ in this case but we suggest that if this is required then the data are not worth the effort.

The final calculation performed is the further integration to obtain the CN for both normalized g(r) functions using the equation below:

$$\text{CN} = 4\pi\rho \int\limits_{r_1}^{r_2} r^2 g(r) dr$$

Mathematically, the CN calculation consists of integrating $r^2 g(r)$ from the minimum before the first peak ($r_1$) to the minimum after the first peak ($r_2$). The code locates the highest peak in $g(r)$, then locates the minima on each side and integrating using the trapz numerical integration function. In contrast to the theoretical definition of the CN (an integration starting from $r = 0$, equation 14 in the main text) the finite lower limit on $r$ has to be included when analysing $g(r)$ originating from FT of real experimental data to avoid including the area of unphysical ripples at very low $r$ in the calculation.

If the unphysical ripples in $g(r)$ caused by the FT process cause the first physically meaningful peak to not be the highest peak, then the procedure will fail. *It is the responsibility of the user to check this by viewing the graphs of the relevant functions produced by the code.*

The CNs calculated from both $g(r)$ functions are outputted in variables that the user can read in the Octave workspace, and the following outputs are saved to file:

- The raw $[g(r) - g_0]$ and normalized $g(r)$ functions obtained with the full $S(Q)$ dataset.
- The normalized g(r) function obtained with the $S(Q)$ dataset over a reduced range in $Q$.

The following figures are created:

- Figure 1. The $S(Q)$ that was provided, over the full range in $Q$.
- Figure 2. The raw $[g(r) - g_0]$ functions obtained from the FT of the full $S(Q)$ dataset and the dataset over a reduced range in $Q$.
- Figure 3. The normalized $g(r)$ functions obtained from the FT of the full $S(Q)$ dataset and the dataset over a reduced range in $Q$.

## 5.2 Fourier transform with Lorch modification function

In this case, when the Fourier transform is performed to obtain $[g(r_0) - g_0]$ equation 6 from the main text is utilized, but with a finite lower limit $Q_{min}$ similarly to the direct FT described above. $Q_{max}$ is the maximum value of Q for which S(Q) data are provided for the FT process, and $\Delta$ is obtained from $Q_{max}$ using $\Delta = \pi/Q_{max}$ as justified in the main text.

**5.3 Fourier transform with Soper-Barney modification function**

In this case, when the Fourier transform is performed to obtain $[g(r_0) - g_0]$ equation 11 from the main text is utilized, but with a finite lower limit $Q_{min}$ similarly to the direct FT described above. $Q_{max}$ is the maximum value of $Q$ for which $S(Q)$ data are provided for the FT process, and $\Delta$ is obtained from $Q_{max}$ using $\Delta = 4.49/Q_{max}$ as justified in the main text.

**6. Code**

It should be possible to copy and paste this code directly into Octave or Matlab. Each code is a separate self-contained script. The code is also available on request in Octave (.m) format.

**6.1 Standard Fourier transform**

## STEP 1: LOAD S(Q) DATA ##

```
clear variables
load Ar_335977.txt;
data_file = Ar_335977;
p = 1.75132E+27; #number density of particles
Qmax = 50; #Q cutoff variables

Q_values = data_file(:,1)*10^10; #Q
S_Q_values = data_file(:,2); #S(Q)
n = 1000; #number of points to be calculated
r_max = 20*10^-10;
r_spacing = r_max/n;
r_values = r_spacing:r_spacing:r_max; #array of r values

Q_value = Q_values(Qmax)*10^-10; #Just for readout, the actual value of Q cutoff in A^-1
Q_values_cutoff = [];
S_Q_values_cutoff = [];

for i = 1:Qmax #input values into Q for cutoff integral
  Q_values_cutoff = [Q_values_cutoff;Q_values(i)];
  S_Q_values_cutoff = [S_Q_values_cutoff;S_Q_values(i)];
end

## STEP 2: FINDING G(R) - G0 ##
raw_g_r = []; #g(r) - g0 with full range
raw_g_r_cutoff = []; #g(r) - g0 with cut-off

for i = 1:length(r_values) #we integrate for all values of r
  #the full-range integral
  F = Q_values.*(S_Q_values-1).*sin(Q_values*r_values(i));
  g = trapz(Q_values,F)*2/pi; #integrate
  gr = g/(4*pi*r_values(i)); #the integral gives 4*pi*r*[g(r) - g0], so divide by 4*pi*r
  raw_g_r = [raw_g_r;gr];
  #the cut-off integral
```

```
    F = Q_values_cutoff.*(S_Q_values_cutoff-1).*sin(Q_values_cutoff*r_values(i));
    g = trapz(Q_values_cutoff,F)*2/pi; #integrate
    gr = g/(4*pi*r_values(i)); #the integral gives 4*pi*r*[g(r) - g0], so divide by 4*pi*r
    raw_g_r_cutoff = [raw_g_r_cutoff;gr];
end

## STEP 3: NORMALIZATION ##

norm_g_r = []; #normalized g(r) for full range
norm_g_r_cutoff = []; #normalized g(r) for cut-off

if (raw_g_r(1) > raw_g_r(length(r_values)))
  disp('Full range g(r) data cannot be normalized');
end

if (raw_g_r_cutoff(1) > raw_g_r_cutoff(length(r_values)))
  disp('Qmax cutoff g(r) data cannot be normalized');
end

for i = 1:length(r_values)
  norm_g_r = [norm_g_r;(raw_g_r(i) - raw_g_r(1))];
  norm_g_r_cutoff = [norm_g_r_cutoff;(raw_g_r_cutoff(i) - raw_g_r_cutoff(1))];
end

for i = 1:length(r_values)
  norm_g_r(i) = norm_g_r(i) / norm_g_r(length(r_values));
  norm_g_r_cutoff(i) = norm_g_r_cutoff(i) / norm_g_r_cutoff(length(r_values));
end

## STEP 4: CO-ORDINATION NUMBER ##

#CN: full range
int_table = [];
Max_Peak = 0;
max_peak_pos = 1; #Position of first Cshell max
coord_max = 1; #Upper limit for CN integration
coord_min = 1; #Lower limit for CN integration
func = norm_g_r;

for i = 2:length(r_values) #locate tallest peak
  if (func(i) > Max_Peak)
    Max_Peak = func(i);
    max_peak_pos = i;
  end
end

#see where minimum is after tallest peak
Lowest = func(max_peak_pos);

for i = (max_peak_pos+1):length(r_values)
  if (Lowest > func(i)) #i.e. gradient is negative
```

```
      Lowest = func(i);
      coord_max = i;
    else
      break
    end
  end
end

#See where minimum is before tallest peak
for i = 2:(max_peak_pos-1)
    if (func(i-1) > func(i)) #So it sets coord_min if there is a negative gradient
      coord_min = i;
    endif
end

for i = coord_min:coord_max #Num should be at the point of the lowest peak
  int_table = [int_table, (r_values(i)^2).*norm_g_r(i)]; #get sets of data points for CN integral
end

area = trapz(r_values(coord_min:coord_max),int_table); #integrate
CN = 4*pi*p*area;

#CN: cutoff range
int_table = [];
Max_Peak = 0;
max_peak_pos = 1; #Position of first Cshell max
coord_max = 1; #Upper limit for CN integration
coord_min = 1; #Lower limit for CN integration
func = norm_g_r_cutoff;

for i = 2:length(r_values) #locate tallest peak
  if (func(i) > Max_Peak)
    Max_Peak = func(i);
    max_peak_pos = i;
  end
end

#see where minimum is after tallest peak
Lowest = func(max_peak_pos);

for i = (max_peak_pos+1):length(r_values)
  if (Lowest > func(i)) #i.e. gradient is negative
    Lowest = func(i);
    coord_max = i;
  else
    break
  end
end

#See where minimum is before tallest peak
for i = 2:(max_peak_pos-1)
    if (func(i-1) > func(i)) #So it sets coord_min if there is a negative gradient
```

```
    coord_min = i;
  endif
end

for i = coord_min:coord_max #Num should be at the point of the lowest peak
  int_table = [int_table, (r_values(i)^2).*norm_g_r_cutoff(i)]; #get sets of data points for CN integral
end

area = trapz(r_values(coord_min:coord_max),int_table); #integrate
CN_cutoff = 4*pi*p*area;

## EXPORTING AND GRAPHS ##
#Make a descending list of r values so we can save them in a text document
r_list = [];

for i = 1:length(r_values)
  r_list = [r_list;r_values(i)*10^10];
end

RDFdata = [r_list norm_g_r];
RDFdata2 = [r_list raw_g_r];
CutOffdata = [r_list norm_g_r_cutoff];
save Std_norm_RDF.txt RDFdata;
save Std_raw_RDF.txt RDFdata2;
save Std_norm_cutoff_RDF.txt CutOffdata;

figure(1)
plot(Q_values,S_Q_values)
xlabel('Q')
ylabel('S(Q)')
title('Structure Factor')
grid on
box on

figure(2)
plot(r_values,raw_g_r,'r',r_values,raw_g_r_cutoff,'b')
legend('Max Q','Q Cutoff')
xlabel('r')
ylabel('G(r) - G0')
title('Radial Distribution Function: Fourier transform')
grid on
box on

figure(3)
plot(r_values,norm_g_r,'r',r_values,norm_g_r_cutoff,'b')
legend('Max Q','Q Cutoff')
xlabel('r')
ylabel('g(r)')
title('Normalised g(r)')
grid on
box on
```

**6.2 Fourier transform with Lorch modification**

## STEP 1: LOAD S(Q) DATA ##

```
clear variables
load Ar_335953.txt;
data_file = Ar_335953;
p = 2.5924E28; #number density of particles
Qmax = 870; #Q cutoff variables

Q_values = data_file(:,1)*10^10; #Q
S_Q_values = data_file(:,2); #S(Q)
n = 1000; #number of points to be calculated
r_max = 20*10^-10;
r_spacing = r_max/n;
r_values = r_spacing:r_spacing:r_max; #array of r values

Q_value = Q_values(Qmax)*10^-10; #Just for readout, the actual value of Q cutoff in A^-1
Q_values_cutoff = [];
S_Q_values_cutoff = [];

delta = pi/Q_values(length(Q_values));
cutoff_delta = pi/Q_values(Qmax);

for i = 1:Qmax #input values into Q for cutoff integral
  Q_values_cutoff = [Q_values_cutoff;Q_values(i)];
  S_Q_values_cutoff = [S_Q_values_cutoff;S_Q_values(i)];
end

## STEP 2: FINDING G(R) - G0 ##
raw_g_r = []; #g(r) - g0 with full range
raw_g_r_cutoff = []; #g(r) - g0 with cut-off

for i = 1:length(r_values) #we integrate for all values of r
  #the full-range integral
  F = Q_values.*(S_Q_values-
1).*sin(Q_values.*r_values(i)).*(sin(Q_values.*delta)./(Q_values.*delta));
  g = trapz(Q_values,F)*2/pi; #integrate
  gr = g/(4*pi*r_values(i)); #the integral gives 4*pi*r*[g(r) - g0], so divide by 4*pi*r
  raw_g_r = [raw_g_r;gr];
  #the cut-off integral
  F = Q_values_cutoff.*(S_Q_values_cutoff-
1).*sin(Q_values_cutoff.*r_values(i)).*(sin(Q_values_cutoff.*cutoff_delta)./(Q_values_cutoff.*cutoff
_delta));
  g = trapz(Q_values_cutoff,F)*2/pi; #integrate
  gr = g/(4*pi*r_values(i)); #the integral gives 4*pi*r*[g(r) - g0], so divide by 4*pi*r
  raw_g_r_cutoff = [raw_g_r_cutoff;gr];
end
```

## STEP 3: NORMALIZATION ##

```
norm_g_r = []; #normalized g(r) for full range
norm_g_r_cutoff = []; #normalized g(r) for cut-off

if (raw_g_r(1) > raw_g_r(length(r_values)))
  disp('Full range g(r) data cannot be normalized');
end

if (raw_g_r_cutoff(1) > raw_g_r_cutoff(length(r_values)))
  disp('Qmax cutoff g(r) data cannot be normalized');
end

for i = 1:length(r_values)
  norm_g_r = [norm_g_r;(raw_g_r(i) - raw_g_r(1))];
  norm_g_r_cutoff = [norm_g_r_cutoff;(raw_g_r_cutoff(i) - raw_g_r_cutoff(1))];
end

for i = 1:length(r_values)
  norm_g_r(i) = norm_g_r(i) / norm_g_r(length(r_values));
  norm_g_r_cutoff(i) = norm_g_r_cutoff(i) / norm_g_r_cutoff(length(r_values));
end
```

## STEP 4: CO-ORDINATION NUMBER ##
```
#CN: full range
int_table = [];
Max_Peak = 0;
max_peak_pos = 1; #Position of first Cshell max
coord_max = 1; #Upper limit for CN integration
coord_min = 1; #Lower limit for CN integration
func = norm_g_r;

for i = 2:length(r_values) #locate tallest peak
  if (func(i) > Max_Peak)
    Max_Peak = func(i);
    max_peak_pos = i;
  end
end

#see where minimum is after tallest peak
Lowest = func(max_peak_pos);

for i = (max_peak_pos+1):length(r_values)
  if (Lowest > func(i)) #i.e. gradient is negative
    Lowest = func(i);
    coord_max = i;
  else
    break
  end
end
```

```
#See where minimum is before tallest peak
for i = 2:(max_peak_pos-1)
   if (func(i-1) > func(i)) #So it sets coord_min if there is a negative gradient
     coord_min = i;
   endif
end

for i = coord_min:coord_max #Num should be at the point of the lowest peak
  int_table = [int_table, (r_values(i)^2).*norm_g_r(i)]; #get sets of data points for CN integral
end

area = trapz(r_values(coord_min:coord_max),int_table); #integrate
CN = 4*pi*p*area;

#CN: cutoff range
int_table = [];
Max_Peak = 0;
max_peak_pos = 1; #Position of first Cshell max
coord_max = 1; #Upper limit for CN integration
coord_min = 1; #Lower limit for CN integration
func = norm_g_r_cutoff;

for i = 2:length(r_values) #locate tallest peak
  if (func(i) > Max_Peak)
    Max_Peak = func(i);
    max_peak_pos = i;
  end
end

#see where minimum is after tallest peak
Lowest = func(max_peak_pos);

for i = (max_peak_pos+1):length(r_values)
  if (Lowest > func(i)) #i.e. gradient is negative
    Lowest = func(i);
    coord_max = i;
  else
    break
  end
end

#See where minimum is before tallest peak
for i = 2:(max_peak_pos-1)
   if (func(i-1) > func(i)) #So it sets coord_min if there is a negative gradient
     coord_min = i;
   endif
end

for i = coord_min:coord_max #Num should be at the point of the lowest peak
  int_table = [int_table, (r_values(i)^2).*norm_g_r_cutoff(i)]; #get sets of data points for CN integral
end
```

```
area = trapz(r_values(coord_min:coord_max),int_table); #integrate
CN_cutoff = 4*pi*p*area;

## EXPORTING AND GRAPHS ##
#Make a descending list of r values so we can save them in a text document
r_list = [];

for i = 1:length(r_values)
  r_list = [r_list;r_values(i)*10^10];
end

RDFdata = [r_list norm_g_r];
RDFdata2 = [r_list raw_g_r];
CutOffdata = [r_list norm_g_r_cutoff];
save Lorch_norm_RDF.txt RDFdata;
save Lorch_raw_RDF.txt RDFdata2;
save Lorch_norm_cutoff_RDF.txt CutOffdata;

figure(1)
plot(Q_values,S_Q_values)
xlabel('Q')
ylabel('S(Q)')
title('Structure Factor')
grid on
box on

figure(2)
plot(r_values,raw_g_r,'r',r_values,raw_g_r_cutoff,'b')
legend('Max Q','Q Cutoff')
xlabel('r')
ylabel('G(r) - G0')
title('Radial Distribution Function: Fourier transform')
grid on
box on

figure(3)
plot(r_values,norm_g_r,'r',r_values,norm_g_r_cutoff,'b')
legend('Max Q','Q Cutoff')
xlabel('r')
ylabel('g(r)')
title('Normalised g(r)')
grid on
box on
```

**6.3 Fourier transform with Soper-Barney modification**

```
## STEP 1: LOAD S(Q) DATA ##

clear variables
```

```
load Ar_335953.txt;
data_file = Ar_335953;
p = 2.5924E28; #number density of particles
Qmax = 870; #Q cutoff variables

Q_values = data_file(:,1)*10^10; #Q
S_Q_values = data_file(:,2); #S(Q)
n = 1000; #number of points to be calculated
r_max = 20*10^-10;
r_spacing = r_max/n;
r_values = r_spacing:r_spacing:r_max; #array of r values

Q_value = Q_values(Qmax)*10^-10; #Just for readout, the actual value of Q cutoff in A^-1
Q_values_cutoff = [];
S_Q_values_cutoff = [];

delta = 4.49./Q_values(length(Q_values));
cutoff_delta = 4.49./Q_values(Qmax);

for i = 1:Qmax #input values into Q for cutoff integral
  Q_values_cutoff = [Q_values_cutoff;Q_values(i)];
  S_Q_values_cutoff = [S_Q_values_cutoff;S_Q_values(i)];
end

## STEP 2: FINDING G(R) - G0 ##
raw_g_r = []; #g(r) - g0 with full range
raw_g_r_cutoff = []; #g(r) - g0 with cut-off

for i = 1:length(r_values) #we integrate for all values of r
  #the full-range integral
  F = Q_values.*(S_Q_values-
1).*sin(Q_values.*r_values(i)).*(3./((Q_values.*delta).^3)).*(sin(Q_values.*delta)-
(Q_values.*delta).*cos(Q_values.*delta));
  g = trapz(Q_values,F)*2/pi; #integrate
  gr = g/(4*pi*r_values(i)); #the integral gives 4*pi*r*[g(r) - g0], so divide by 4*pi*r
  raw_g_r = [raw_g_r;gr];
  #the cut-off integral
  F = Q_values_cutoff.*(S_Q_values_cutoff-
1).*sin(Q_values_cutoff.*r_values(i)).*(3./((Q_values_cutoff.*cutoff_delta).^3)).*(sin(Q_values_cuto
ff.*cutoff_delta)-(Q_values_cutoff.*cutoff_delta).*cos(Q_values_cutoff.*cutoff_delta));
  g = trapz(Q_values_cutoff,F)*2/pi; #integrate
  gr = g/(4*pi*r_values(i)); #the integral gives 4*pi*r*[g(r) - g0], so divide by 4*pi*r
  raw_g_r_cutoff = [raw_g_r_cutoff;gr];
end

## STEP 3: NORMALIZATION ##

norm_g_r = []; #normalized g(r) for full range
norm_g_r_cutoff = []; #normalized g(r) for cut-off

if (raw_g_r(1) > raw_g_r(length(r_values)))
```

```
  disp('Full range g(r) data cannot be normalized');
end

if (raw_g_r_cutoff(1) > raw_g_r_cutoff(length(r_values)))
  disp('Qmax cutoff g(r) data cannot be normalized');
end

for i = 1:length(r_values)
  norm_g_r = [norm_g_r;(raw_g_r(i) - raw_g_r(1))];
  norm_g_r_cutoff = [norm_g_r_cutoff;(raw_g_r_cutoff(i) - raw_g_r_cutoff(1))];
end

for i = 1:length(r_values)
  norm_g_r(i) = norm_g_r(i) / norm_g_r(length(r_values));
  norm_g_r_cutoff(i) = norm_g_r_cutoff(i) / norm_g_r_cutoff(length(r_values));
end
```

## STEP 4: CO-ORDINATION NUMBER ##

```
#CN: full range
int_table = [];
Max_Peak = 0;
max_peak_pos = 1; #Position of first Cshell max
coord_max = 1; #Upper limit for CN integration
coord_min = 1; #Lower limit for CN integration
func = norm_g_r;

for i = 2:length(r_values) #locate tallest peak
  if (func(i) > Max_Peak)
    Max_Peak = func(i);
    max_peak_pos = i;
  end
end

#see where minimum is after tallest peak
Lowest = func(max_peak_pos);

for i = (max_peak_pos+1):length(r_values)
  if (Lowest > func(i)) #i.e. gradient is negative
    Lowest = func(i);
    coord_max = i;
  else
    break
  end
end

#See where minimum is before tallest peak
for i = 2:(max_peak_pos-1)
    if (func(i-1) > func(i)) #So it sets coord_min if there is a negative gradient
      coord_min = i;
    endif
```

```
end

for i = coord_min:coord_max #Num should be at the point of the lowest peak
  int_table = [int_table, (r_values(i)^2).*norm_g_r(i)]; #get sets of data points for CN integral
end

area = trapz(r_values(coord_min:coord_max),int_table); #integrate
CN = 4*pi*p*area;

#CN: cutoff range
int_table = [];
Max_Peak = 0;
max_peak_pos = 1; #Position of first Cshell max
coord_max = 1; #Upper limit for CN integration
coord_min = 1; #Lower limit for CN integration
func = norm_g_r_cutoff;

for i = 2:length(r_values) #locate tallest peak
  if (func(i) > Max_Peak)
    Max_Peak = func(i);
    max_peak_pos = i;
  end
end

#see where minimum is after tallest peak
Lowest = func(max_peak_pos);

for i = (max_peak_pos+1):length(r_values)
  if (Lowest > func(i)) #i.e. gradient is negative
    Lowest = func(i);
    coord_max = i;
  else
    break
  end
end

#See where minimum is before tallest peak
for i = 2:(max_peak_pos-1)
  if (func(i-1) > func(i)) #So it sets coord_min if there is a negative gradient
    coord_min = i;
  endif
end

for i = coord_min:coord_max #Num should be at the point of the lowest peak
  int_table = [int_table, (r_values(i)^2).*norm_g_r_cutoff(i)]; #get sets of data points for CN integral
end

area = trapz(r_values(coord_min:coord_max),int_table); #integrate
CN_cutoff = 4*pi*p*area;

## EXPORTING AND GRAPHS ##
```

```
#Make a descending list of r values so we can save them in a text document
r_list = [];

for i = 1:length(r_values)
  r_list = [r_list;r_values(i)*10^10];
end

RDFdata = [r_list norm_g_r];
RDFdata2 = [r_list raw_g_r];
CutOffdata = [r_list norm_g_r_cutoff];
save Soper_norm_RDF.txt RDFdata;
save Soper_raw_RDF.txt RDFdata2;
save Soper_norm_cutoff_RDF.txt CutOffdata;

figure(1)
plot(Q_values,S_Q_values)
xlabel('Q')
ylabel('S(Q)')
title('Structure Factor')
grid on
box on

figure(2)
plot(r_values,raw_g_r,'r',r_values,raw_g_r_cutoff,'b')
legend('Max Q','Q Cutoff')
xlabel('r')
ylabel('G(r) - G0')
title('Radial Distribution Function: Fourier transform')
grid on
box on

figure(3)
plot(r_values,norm_g_r,'r',r_values,norm_g_r_cutoff,'b')
legend('Max Q','Q Cutoff')
xlabel('r')
ylabel('g(r)')
title('Normalised g(r)')
grid on
box on
```